 \documentclass[final,authoryear,5p,times,twocolumn]{elsarticle}
 \usepackage[T1]{fontenc}
\usepackage[latin1]{inputenc}
\usepackage[english]{babel}

\usepackage[mediumqspace,mediumspace,amssymb,Gray]{SIunits}

\usepackage{amsmath}
\usepackage{amstext}
\usepackage{latexsym}
\usepackage{amsfonts}
\usepackage{mathrsfs}
\usepackage{bm}

\usepackage{url}
\usepackage{multirow}
\usepackage{dcolumn}
\newcolumntype{d}{D{.}{.}{-1}}
\newcolumntype{f}[1]{D{.}{.}{#1}}

\usepackage{bm}
\usepackage[colorlinks=true,linkcolor=blue]{hyperref}%

\usepackage{color}

\usepackage[normalem]{ulem}

\biboptions{sort}

\def\asc2#1{{\color{red}{#1}}} 
   
\def\asq#1{ {\color{red} {..(??. #1 .??)..} } }         
\def\asq2#1{ {\color{red} {? #1 ?} } }         
\journal{Radiation Physics and Chemistry}

\begin{document}
\begin{frontmatter}


\title{A vacuum double-crystal  spectrometer for reference-free highly charged ions X-ray spectroscopy}

%
%
%

\author{P.\ Amaro\fnref{cfa,lkb,heidel}}
\address[cfa]{Centro de F\'isica At\'omica, CFA, Departamento de F\'isica, Faculdade de Ci\^encias e Tecnologia, FCT, Universidade Nova de Lisboa, 2829-516 Caparica, Portugal}
\address[lkb]{Laboratoire Kastler Brossel, \'Ecole Normale Sup\'erieure, CNRS, Universit\'e P. et M. Curie -- Paris 6, Case 74; 4, place Jussieu, 75252 Paris CEDEX 05, France}
\address[heidel]{Present address: Physikalisches Institut, Heidelberg University, D-69120 Heidelberg, Germany}

\author{Csilla I. Szabo\fnref{lkb}}

\author{Sophie Schlesser\fnref{lkb,kvi}}
\address[kvi]{Present address: KVI, Theory Group, University of Groningen, 9747 AA Groningen, The Netherlands}

\author{A.\ Gumberidze\fnref{lkb,emmi,fias}}
\address[emmi]{ExtreMe Matter Institute EMMI and Research Division, GSI Helmholtzzentrum für Schwerionenforschung, D-64291 Darmstadt, Germany}
\address[fias]{FIAS Frankfurt Institute for Advanced Studies, D-60438 Frankfurt am Main, Germany}

\author{E.G. Kessler, Jr\fnref{NIST}}
\address[NIST]{National Institute of Standards and Technology NIST, 100 Bureau Drive, Gaithersburg, MD 20899, USA}

\author{A. Henins\fnref{NIST}}

\author{E.O. Le Bigot\fnref{lkb}}

\author{M. Trassinelli\fnref{INSP}} 
\address[INSP]{Institut des Nanosciences de Paris, CNRS, Universit\'e P. et M. Curie -- Paris 6, 4, place Jussieu, 75252 Paris CEDEX 05, France}

\author{Jean-Michel Isac\fnref{lkb}}

\author{Pascal Travers\fnref{lkb}}

\author{M.\ Guerra\fnref{cfa}}

\author{J.\ P.\ Santos\fnref{cfa}}

\author{Paul Indelicato\fnref{lkb}\corref{cor1}}
\ead{paul.indelicato@lkb.ens.fr}
\cortext[cor1]{Corresponding Author}

\date{July 29th, 2012}

%

\begin{abstract}
We have built a vacuum double crystal spectrometer, which  coupled to an electron-cyclotron resonance ion source, allows to measure low-energy x-ray transitions in highly-charged ions with accuracies of the order of a few parts per million. We describe in detail the instrument and its performances. Furthermore, we present a few spectra of  transitions in Ar$^{14+}$ , Ar$^{15+}$ and Ar$^{16+}$. We have developed an \emph{ab initio} simulation code that allows us to obtain accurate line profiles. It can reproduce experimental spectra with unprecedented accuracy. The quality of the profiles allows the direct determination of line width. 
\end{abstract}

%
%


\end{frontmatter}

\

\section{Introduction}
\label{sec:intro}
The measurement of x-ray transition energies of highly-charged ions (HCI) is one of the main methods to test bound-state Quantum Electrodynamics (BSQED)  effects in strong fields. Highly-charged ions can be created, e.g., using high-energy accelerators, Electron Beam Ion Traps (EBIT), or Electron-Cyclotron Resonance Ion sources (ECRIS). Transitions between excited states and the $n=1$ ground state in few-electron atoms or ions have been measured in a number of elements ranging from hydrogen to uranium. For medium atomic number elements, relevant to x-ray reflection Bragg spectrometry (transition energies in the \unit{2}{keV} to \unit{15}{keV} range), accuracies in the few tens of parts per million range have been obtained. Beam-foil spectroscopy has been employed to provide measurements in hydrogenlike and heliumlike ions like phosphorus, sulfur, argon \cite{bmic1983} (\unit{80}{ppm}), iron \cite{btim1983,btmd1984}  (\unit{90}{ppm}) \cite{igtb1986}  (\unit{25}{ppm}), germanium \cite{clsd2009} and krypton \cite{tbil1985,itbl1986}. The main limitation to obtain high accuracy in Beam-foil spectrometry is due to the Doppler effect. Correcting for the Doppler shift  requires precise determination of the ions speed and angle of observation of the x rays emitted in flight. To get rid of this uncertainty, argon was also studied by x-ray spectrometry of recoil ions with an accuracy of \unit{5}{ppm} \cite{dbf1984,bdfl1985} relative to an x-ray standard.  The uncertainty then was due to the presence of satellite lines associated with electron capture in the target gas. Another method to reduce the Doppler effect was to decelerate the beam after stripping it at high energy by capturing electrons from a gas cell, in the so called ``accel-decel'' method. Hydrogenlike nickel was studied by this method \cite{bifl1991} with an accuracy of \unit{13}{ppm}.

More recently, devices like Electron Beam Ion Traps (EBIT), which create ions at low-energy, thus reducing considerably the Doppler effect have been used in a number of experiments. Transition energies have been measured in hydrogenlike chlorine \cite{bbkl2007}, and heliumlike argon  \cite{bbkl2007} and vanadium  \cite{cphs2000}. In Ref.   \cite{bbkl2007},  the hydrogenlike chlorine Lyman $\alpha$ lines  are measured without the use of x-ray reference lines, with an accuracy of \unit{10}{ppm}. The accuracy was later improved to   \unit{1.5}{ppm} without external reference \cite{kbbl2012}. This work uses a single Bragg crystal coupled to a CCD camera, which can be positioned very accurately with a laser beam reflected by the same crystal as the x~rays.  The reason to avoid the use of x-ray reference lines is the following. Present day x-ray standards, as can be found in  \cite{dkib2003}, even though they are known with accuracies in the ppm range, are based on neutral elements with a K hole created by electron bombardment or photoionization. The shape and peak position of those lines depend on many factors like the excitation energy (see., e.g.,  Refs.  \cite{dcl1982,dlch1983,dghk1996}), the chemical composition and the surface contamination of the sample. Physical effects like shake-off, Auger and Coster-Kronig effects lead to multivacancies, that  distort and broaden the line shape. Examples of the complex structure of  K$\alpha$ lines in transition elements can be found, e.g., in Refs.  \cite{dhhw1995,hfdh1997,dfhh2004}. It is thus very difficult to use these standard lines with their quoted accuracy. Therefore, it was recently proposed to use either exotic atoms  \cite{agis2003} or highly charged ions   \cite{abbd2003} to provide reliable, reproducible, narrow, x-ray standard lines. 

The first observation of strong x-ray lines of highly charged argon ions (up to He-like)  in an ECRIS was made in 2000 \cite{dkgb2000}. This experiment lead to the description of the mechanisms at work on the production of the different lines in the plasma \cite{mcsi2001,cmps2001}. Since then, several experiments have been performed at the Paul Scherrer Institute (PSI), using a spherically curved crystal spectrometer and an ECRIS, \cite{abgg2005,ibtg2006,ibcg2007,tbcg2007,lbcg2009} leading to improved understanding of the ECRIS plasmas for Sulfur, Chlorine and Argon \cite{smci2008,mmcs2009,scmm2010,smcm2011}.
Such lines can  be used, e.g.,  to characterize x-ray spectrometers response functions\cite{abgg2005}. Yet specific techniques are required to measure their energy without the need for reference lines. The technique of  Ref.   \cite{bbkl2007}, using a single flat crystal, is well adapted to the EBIT, which provides a very narrow (\unit{\approx 100}{\mu m}), but rather weak x-ray source. The ECRIS plasmas have been shown to be very intense sources of x~rays, but have diameters of a few cm. They are thus better adapted to spectrometers that can use an extended source. At low energies, cylindrically- or spherically-bent crystal spectrometers and double-crystal spectrometers (DCS) can be used, but only the latter can provide high-accuracy, reference-free measurements. 

Precision spectroscopy with double-crystal x-ray spectrometers has a long history. The first DCS was conceived and employed independently by  \citet{com1917},   \citet{bjb1921} and  \citet{wak1922} to measure absolute integrated reflections of crystals.   \citet{das1921} used the DCS to study the width of the reflection curve. These experiments showed that the DCS was an instrument of high precision and high resolving power. They were followed by several others  (see, e.g., Refs.  \cite{aaw1930,all1932}), and were instrumental in establishing the dynamical diffraction theory of   \citet{dar1914,dar1914a} and  \citet{pri1930}. The properties of the dispersive mode to reach high-resolution was found by  \citet{dap1927,dap1928}. The DCS was then used to  obtain the  K-line widths of some elements \cite{aaw1930,all1933}.  A theoretical description of the instrument was provided by  \citet{sch1928}. A detailed technical  description was given by  \citet{com1931}  and   \citet{waa1929}.  \citet{wil1932} introduced the vertical divergence correction in 1932, allowing for an improved accuracy for energy measurements.  \citet{bea1931} provided an absolute measurement of copper and chromium K lines with the use of ruled gratings and calcite crystals in a DCS and deduced a value for the calcite lattice spacing, \cite{bea1931a} leading the way to absolute x-ray wavelength measurements \cite{bea1932}. Detailed description of the instrument can be found in classic textbooks \cite{caa1935,jam1948}.

In a DCS, the first crystal, which is kept at a fixed angle, acts as a collimator, defining the direction and the energy of the incoming x-ray beam, which is analyzed by the second crystal. A first peak is obtained by scanning the second crystal  angle when the two crystals are parallel (non-dispersive mode).  \citet{wak1922} were the first to show the absence of dispersion in the parallel mode. The peak shape depends only on the reflection profile of the crystals and provides the response function of the instrument. A second peak is obtained when both crystals deflect the beam in the same direction (dispersive mode). The peak shape is then a convolution of the line shape and of the instrument response function. The position of the first crystal is the same in both modes. The difference in angle settings of the second crystal between the non-dispersive and the dispersive modes is directly connected to the Bragg angle. The DCS can be used in reflection (low-energy x~rays), in which case the energy that is being measured depends only on the Bragg angle, on the crystals lattice spacing $d$, on the crystal index of refraction  and on the geometry (distance between the entrance and exit slits and height and width of the slits) of the instrument. In this case the reflecting planes are parallel to the surface of the crystal. In transmission (high-energy), there is no index of refraction correction, and the reflecting planes are perpendicular to the surface. The DCS in both modes was used for many years to measure x-ray energies relative to a standard lines, as the crystal lattice spacing was not known. This changed dramatically when high-quality Si and Ge high-purity single crystals became available, which were needed for the fabrication of transistors. Interferometric methods were then developed to do direct measurements of the lattice spacing in term of laser wavelength, with accuracies below $10^{-8}$ \cite{bah1965,har1968,dah1973,bdel1981,oat1984,fmm2008,mmk2009,mmkf2009,mmk2009}. The DCS became then a way to do reference-free measurements of x-ray wavelengths, using well-measured and characterized crystals as transfer standards.  \citet{des1967} designed a vacuum DCS  intended for low-energy x~rays measurements, with high-precision angular encoders and rotating tables \cite{des1967}. A high-precision transmission instrument was constructed for high-energy x and $\gamma$ ~rays, with angular interferometers able to measure angles to a fraction of milliarcseconds. This instrument allowed to resolve inconsistencies between different determination of  x and $\gamma$ ~ray wavelengths \cite{kdh1979,dksh1980}. DCSs have been used to measure K lines of light elements like magnesium \cite{sdmp1994}, copper \cite{dhhw1995} and heavy elements like tungsten  \cite{kdh1979} and from silver to uranium \cite{kdgs1982}. A complete tabulation of all available x-ray standards can be found in Ref.  \cite{dkib2003}.


The purpose of this paper is twofold: first we describe a vacuum DCS for low energy x~rays, adapted to the use of an ECRIS plasma as x-ray source, which provides specific constraints as the ECRIS cannot be moved, contrary to an x-ray tube, to be set to the correct position for a given Bragg angle. Second we discuss the performance and properties of the system of a DCS coupled to an ECRIS and describe the \emph{ab initio} simulation code that we develop to reproduce and analyze experimental spectra.

We describe a method to provide \emph{absolute} measurements (without external reference) of line energies
of inner-shell transitions in highly charged ions with an \emph{accuracy}
unavailable until now. Reaching an accuracy of a few parts per million
(ppm) in this context, can probe and test QED (Quantum ElectroDynamic) effects such as two-loop self energy
corrections and provide new, more reliable x-ray standards in the few keV energy region \cite{agis2003, abgg2005}.


%
This article is organized as follows: in Sec. \ref{sec:setup} we describe the technical features of the ECRIS and the
DCS as well as their alignment. We also describe the measurement of the lattice spacing of the crystals that are used by the DCS. In Sec. \ref{sec:theoryDCS} we give a brief review of the Monte-Carlo simulation used in this work. Furthermore, we discuss the experimental procedure in Sec. \ref{sec:data_an} and how one can assess the accuracy  of measurements performed with a DCS in Sec. \ref{sec:uncertain}. In Sec. \ref{sec:results} we present an example of measurements and experimental tests performed with this experimental setup. 
The conclusions are presented in Sec. \ref{sec:concl}.

%
\section{Experimental setup}
\label{sec:setup}%
%
\subsection{ECRIS}
%
%
An electron-cyclotron resonance ion source is a device built around a minimum-$B$ structure, designed to trap hot electrons. The structure is composed of a magnetic bottle for longitudinal trapping, that can be made of coils or permanent magnets. An ion beam can be extracted along the symmetry axis of the bottle by applying high voltage. Transverse trapping is usually performed with a multlipole magnet, e.g., in our case a hexapole magnet. Microwaves are injected in a plasma chamber inside this structure, at a frequency resonant with the electrons cyclotron frequency on a constant $\mid \boldsymbol{B}\mid$ surface, which resembles an ellipsoid. The electric field of the microwaves can then accelerate electrons to very high energies. A gas or vapor is injected inside the plasma chamber and the atoms are then ionized and trapped in the space charge of the electrons, which have a density in the order of \unit{10^{11}}{cm^{-3}}. The plasma is subject to an electric field through a \emph{polarization electrode} which helps to optimize the ion production. A general description of an ECRIS can be found in, e.g.,  Ref.  \cite{gel1996}.

\begin{figure}
[htb]
\centering
\includegraphics[clip=true,width=\columnwidth]{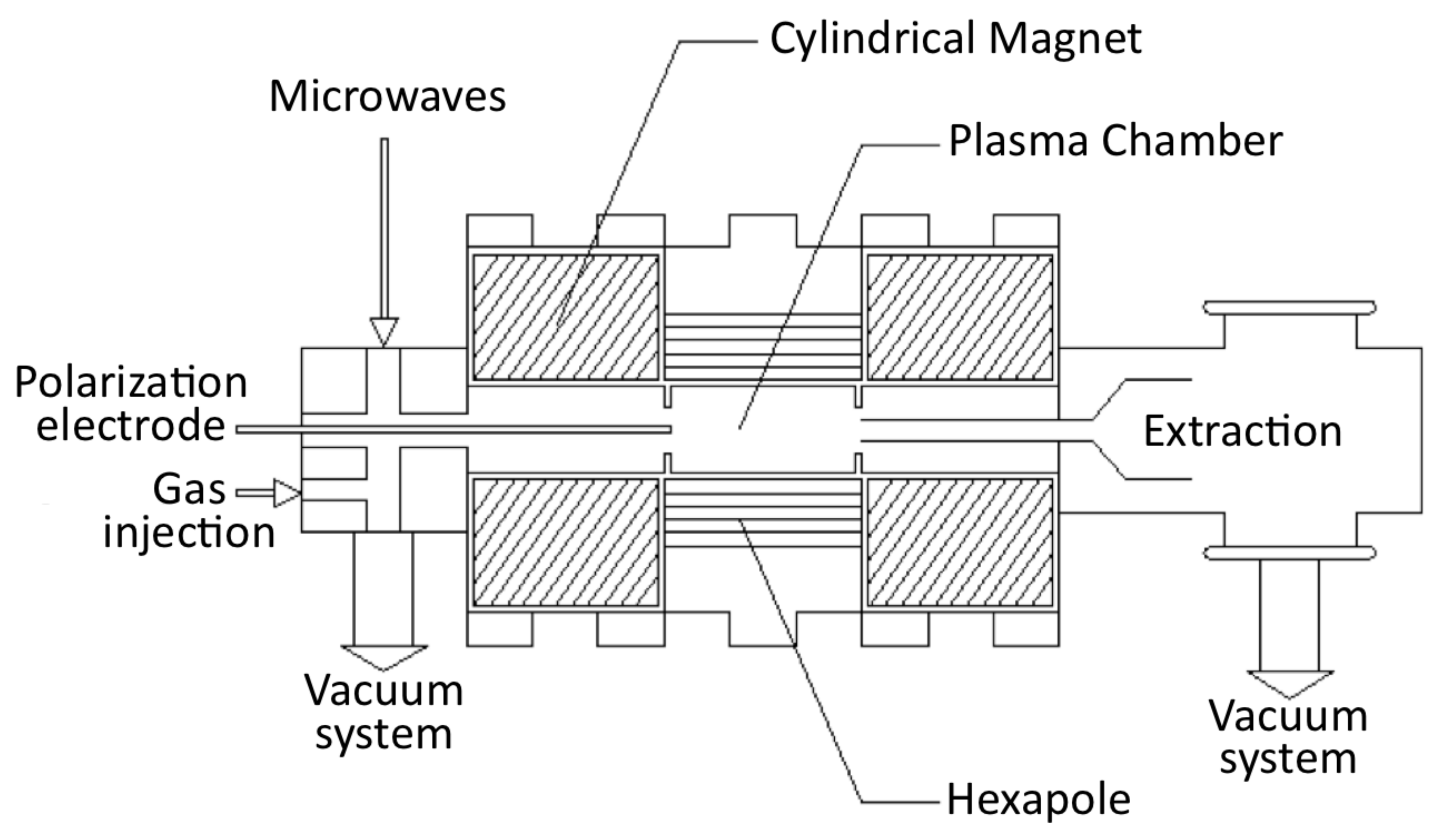}
\caption{Principle scheme of the SIMPA ECRIS}
\label{fig:simpa}
\end{figure}

The Source d'Ions Multicharg\'{e}s de Paris (SIMPA), is a \textquotedblleft supernanogan\textquotedblright\ ECRIS constructed by the Pantechnik Company \cite{bbck2000}. The magnetic structure is made of permanent magnets, with field strength up to \unit{1.3}{T} at contact. The microwave frequency is \unit{14.5}{GHz}, produced by a \unit{2}{kW} klystron. This source has been jointly operated by the Laboratoire Kastler Brossel (LKB) and the Institut des NanoSciences de Paris (INSP) since 2004. Numerous projects that use the extracted beam and the x-ray radiation of the ECRIS plasma have been started in atomic, plasma and surface physics \cite{gasi2009,gtas2010}. SIMPA has been modified to allow for observation of the plasma though the polarization electrode. A sketch of the SIMPA ECRIS is presented in Fig. \ref{fig:simpa}. The plasma in SIMPA has roughly a spherical shape, with a diameter of \unit{\approx 3.3}{cm}. 

The source has been fully characterized (electronic and ionic densities, electronic temperature, x-ray production) using x rays and extracted ion beams \cite{gtas2010}. One consequence of the ion creation and excited level population mechanisms described in Refs.  \cite{mcsi2001,cmps2001,smci2008,mmcs2009,scmm2010,smcm2011} is that the $1s 2s \,^3S_1$ level in He-like ions is strongly populated. It is created by ionization of the Li-like ion ground state $1s^2 2s$. Other excited levels of He-like ions, populated by excitation of the $1s^2$ ground state of He-like ions, or by ionization and excitation mechanisms, are much less populated. This leads to the observation of a very strong $1s 2s \,^3S_1 \to 1s^2\,^1S_0$ M1 transition, which is very forbidden, having a radiative width of only \unit{10^{-7}}{eV}. At the same time, the ions in the ECRIS are rather cold.  They are trapped in a potential created by the space charge of the electrons, with a density of \unit{10^{11}}{cm^{-3}}, which corresponds to a potential depth lower than \unit{1}{eV}. From this a Doppler broadening of \unit{\approx 100}{meV} can be inferred. In contrast, HCI in an EBIT have  higher temperatures due to a deeper trap. The ion temperature in an EBIT was measured recently in Heidelberg and was found to be \unit{6.9}{eV} \cite{scbg2007}  after evaporative cooling. This process further reduces the number of ions that can be used for spectroscopy. A detailed study of this cooling technique was recently performed in a  laser spectroscopy experiment on Ar$^{13+}$ \cite{mkbc2011}.

Observation and measurement of the $1s 2s \,^3S_1 \to 1s^2\,^1S_0$ M1 transition offers a \emph{unique} opportunity to fully characterize a spectrometer. For argon, for example, the Doppler-broadened M1 transition is roughly 8 times narrower than the width of the K$\alpha$ transitions in core-excited argon (\unit{0.79}{eV}) \cite{cap2001}. In the case of the DCS, for which the response function can be calculated from first principles, we can thus compare quantitatively the experimental profile and the simulated one and check the quality of the crystals.

\begin{figure*}
[htb]
\centering
\includegraphics[clip=true,width=\textwidth]{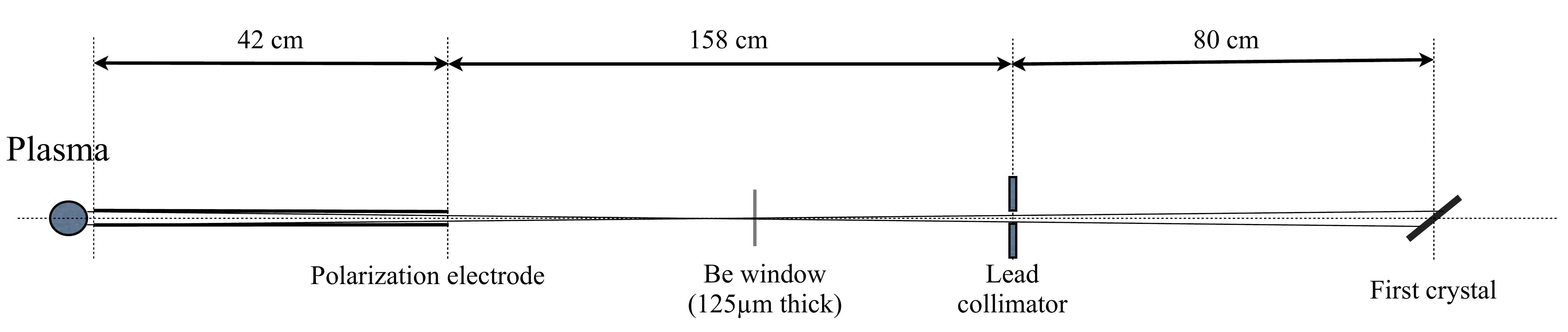}
\caption{Geometrical arrangement of the DCS, polarization electrode, lead collimator and of the SIMPA ECRIS plasma. The inner diameter of the polarization electrode is \unit{12}{mm}.}
\label{fig:simpa-DCS-geom}
\end{figure*}

\begin{figure*}
[htb]
\centering
\includegraphics[clip=true,width=\textwidth]{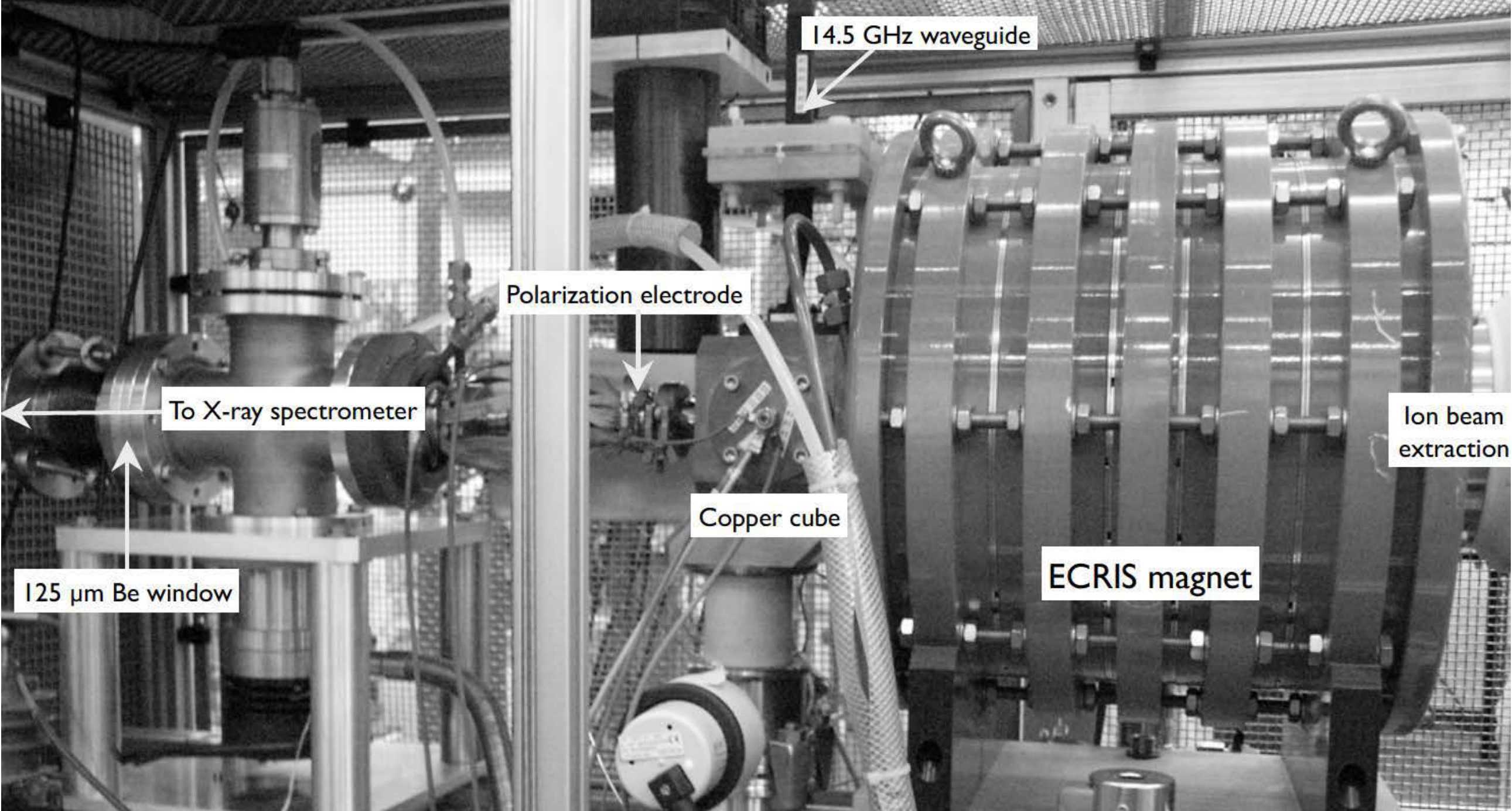}
\caption{The SIMPA ECRIS and its connection to the spectrometer}
\label{fig:simpa-photo-connexion}
\end{figure*}

The geometry of the SIMPA ECRIS has some influence on the positioning of the spectrometer. Figure \ref{fig:simpa-DCS-geom} shows the distances between the plasma, the different parts and the first crystal. A picture of the installation from the source side is shown in Fig. \ref{fig:simpa-photo-connexion}. The installation of a collimator is needed to reduce the background, due, e.g., to x rays that get to the crystal without passing inside the polarization electrode. Because of the collimator and polarization electrode, the x-ray beam that hits the first crystal has an angular aperture of \unit{\pm 6}{Deg}. The Be window, which isolates the vacuum of the source from the primary vacuum in the spectrometer has a transmission varying from 61\% at \unit{2991}{eV} to 65\% at \unit{3135}{eV}, an energy  range that corresponds to the observation of the $1s 2s^2 2p\, ^1P_1\to 1s^2 2s^2 \,^1S_0$ transition in Be-like argon to the  $1s  2p\, ^1P_1\to 1s^2  \,^1S_0$ transition in He-like argon.

All the experiments to date with this setup were performed with few-electron argon ion x~rays. The microwave power injected in the source was between \unit{250}{W} and \unit{350}{W}. A support gas, oxygen, was injected simultaneously with argon, to provide electrons. The pressure, measured at the injection side of the ECRIS (see Fig. \ref{fig:simpa-photo-connexion}), was between \unit{3\times 10^{-5}}{mbars}  and \unit{8\times 10^{-5}}{mbars}. A quadrupole mass spectrometer, positioned on the extraction side of the ECRIS provides  information on the exact composition of the gas in the source, to improve the reproducibility of the ECRIS tuning. 

\subsection{Double Crystal Spectrometer}
\label{subsec:dcs}
%
%

\begin{figure}
[htb]
\centering
\includegraphics[clip=true,width=\columnwidth]{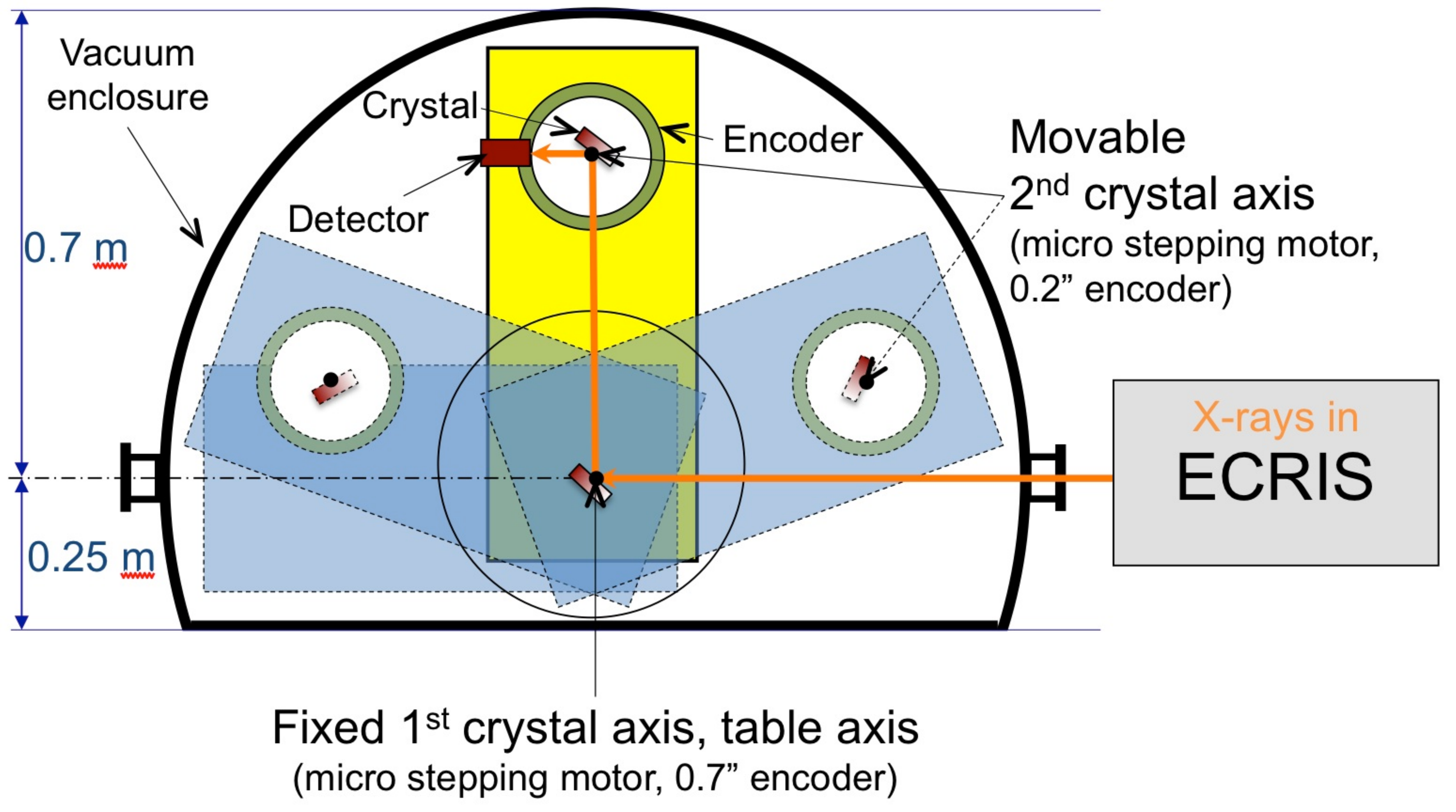}
\caption{Scheme of the DCS.}
\label{fig:SCE_DCS}
\end{figure}

The most characteristic aspect of the DCS at SIMPA, compared to other
double crystal instruments, such as the one located at NIST (National
Institute of Standard and Technology)  \cite{des1967}, is that both crystal axes are mounted on a single support table that rotates around the first crystal axis (Fig. \ref{fig:SCE_DCS}). In other DCSs, the crystals are fixed on a steady platform with the x-ray source having a rotation axis concentric with the first crystal axis \cite{des1967}. The x-ray source can then be rotated so that the x-rays impinging on the first crystal  meet the Bragg condition and are refracted toward the second crystal.

\begin{figure*}
[htb]
\centering
\includegraphics[clip=true,width=\textwidth]{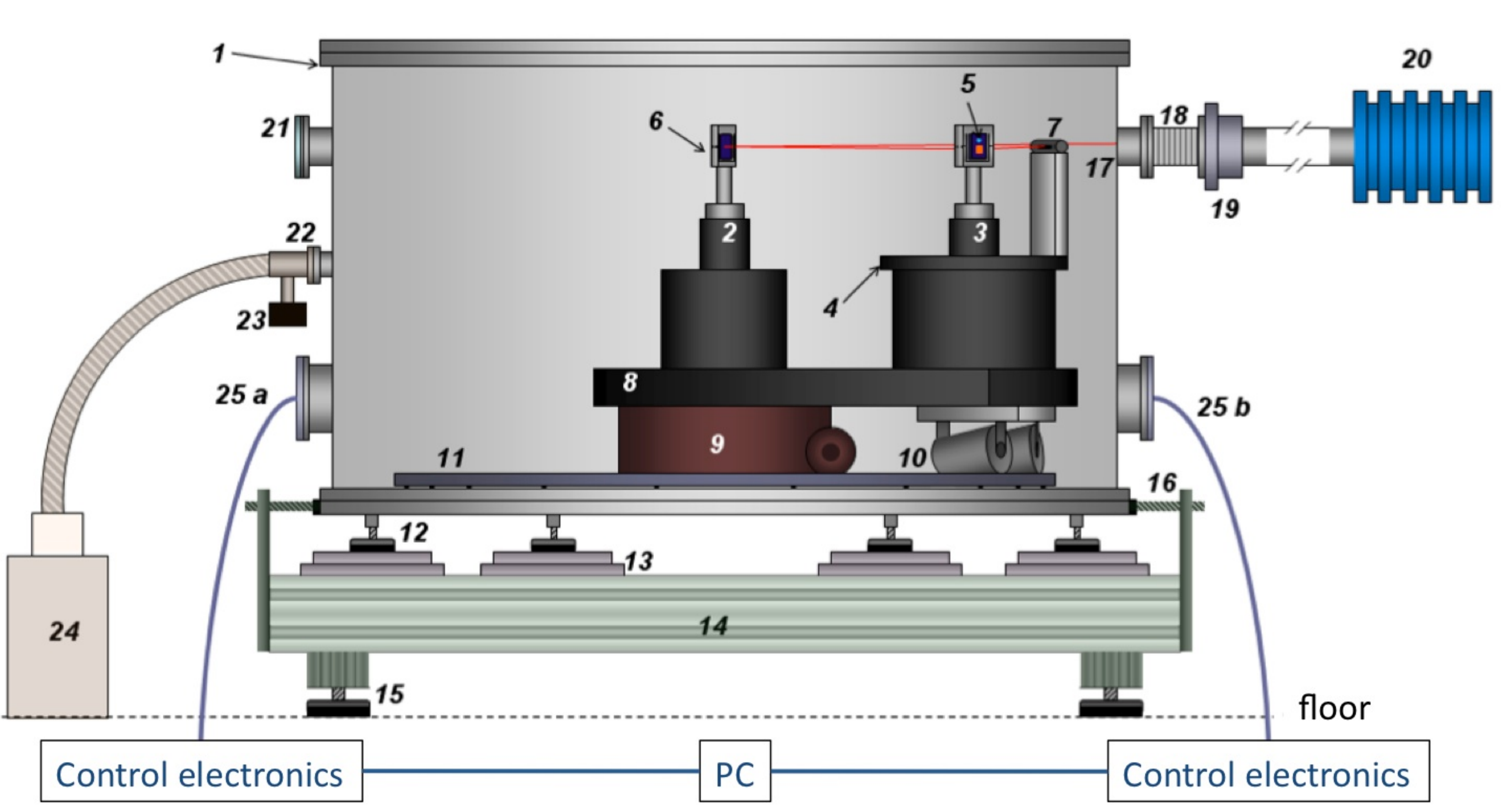}
\caption{Spectrometer setup:  1) vacuum chamber;  2) axis \#1 (first crystal support with rotation stage and angular encoder);  3) axis \#2 (second crystal support, rotation stage, encoder); 4) x-ray  detector rotation stage ;  5)  crystal on second axis;  6) first axis crystal holder;  7) x-ray detector; 8) spectrometer table;  9) spectrometer table rotation stage; 10)  conic wheels;  11) tracks for wheels; 12) vacuum chamber anti-vibration feet with vertical positioning; 13) translation stages; 14) spectrometer support table; 15) anti-vibration feet  with vertical positioning; 16) positioning screws;  17) x-ray entrance; 18) bellows;  19) Be window;  20) SIMPA ECRIS;  21) optical window;  22) bellows connection to vacuum pump; 23) pressure gauge and valve;  24) primary vacuum pump;  25) a and b flanges equipped with feedthroughs for cables and cooling water}
\label{fig:setup_DCS}
\end{figure*}

In this experiment, the x-ray source is a massive, complicated device with several tons of fixed components (vacuum system, magnets, beam line), which makes its rotation impossible. In the design of our DCS,  a heavy table supports both crystal axes, and can be rotated to adjust the instrument to an arbitrary energy range. Figure \ref{fig:setup_DCS} shows an overall view of the spectrometer with all the major components. Both crystals supports are mounted on a single horizontal table (Fig.~\ref{fig:setup_DCS}, 8), \unit{6}{cm} thick and weighing \unit{\approx 200}{kg}. Both supports are built so that the crystals  rotate around a vertical axis (Fig.~\ref{fig:setup_DCS}, 2 and 3) passing through the center of the front surface of each crystal.  The spectrometer table and the rotating table supports are made of a special alloy, LK3, (0.4\% C, 1.8\% Cr, 1\% Al, 0.25\% Mo), chosen for its long-term stability. We used  material that was forged at  a temperature of \unit{1100}{°C} and a stabilized at \unit{900}{°C}  for \unit{48}{h}. After machining, the different parts have been submitted to a stabilization annealing at \unit{825}{°C} for \unit{24}{h} to release  strains in the material. The parts have then been finished by grinding the different surfaces to \unit{2}{\mu m} accuracy to insure excellent parallelism of the two axes. The first crystal support weighs \unit{63}{kg} and the second one \unit{80}{kg}. 

The spectrometer table is mounted on a heavy-duty indexing table (Fig.~\ref{fig:setup_DCS},  9), able to support the weight of the spectrometer assembly (\unit{\approx 360}{kg}), and  rotate it to an arbitrary angle. The assembly rotates around the same vertical axis as the first crystal axis. The indexing table is directly fixed to the lower flange of the vacuum chamber, on a surface that has been precisely machined. Because the spectrometer table is not centered on the indexing table, it is supported by a pair of conical wheels  with precision ball-bearings. The conical part of the wheels has been ground to provide excellent contact. The conical wheels roll on a metallic track  (Fig.~\ref{fig:setup_DCS},  11), resting on the lower vacuum chamber flange, with a system of adjustment screws.  Both the track and the cone surfaces have been hardened. The wheel positions can be adjusted to compensate for the table weight. The vacuum chamber weighs more than \unit{1000}{kg}. It is placed on a support table with adjustable anti-vibration feet. The whole chamber can be moved for alignment with translation stages  (Fig.~\ref{fig:setup_DCS},  13) made of two flat greased metallic pieces. Four screws  (Fig.~\ref{fig:setup_DCS} , 16)  allow to precisely position the chamber during alignment. The support table itself  (Fig.~\ref{fig:setup_DCS}, 14)  rests on the ground with adjustable anti-vibration feet. The chamber can be pumped down to a primary vacuum of \unit{10^{-2}}{mbar} that reduces the absorption of the low energy x rays (around 3 keV), while  being in a range of the Paschen curve for air  where the detector high-voltage (\unit{2}{kV}) does not spark.

The rotation of the crystals is performed with precision stepping motors powered by a three-axis micro-stepping controller Newport ESP301-3G, able to perform rotations as small as 0.017". A Huber model 410 rotation stage is used for the first crystal, a Newport RV80PP for the second crystal and a Newport RV240PP for the detector. The angle of the first crystal is measured with a Heidenhain ROD800 encoder with a sensitivity  of 0.01". Absolute angle is known with 0.5" accuracy over a full turn. The electronic control system uses the digital signal provided by the encoder to maintain the position of the crystal to the set angle over long periods of time. When the angle drifts too far away from the set position, the system stops counting x~rays until the feedback control brings the angle back to the set position. For the data analysis, we use the average first crystal angle, and the standard deviation is used to define the uncertainty. The measured value for the first axis angle standard deviation ranges between 0.014" and 0.065". These positions fluctuations have  a very small contribution to the total error budget. The second crystal angle is measured to a precision of 0.2'' with a Heindenhain RON905 encoder, using a Heidenhain AWE1024 controller for data processing. During data acquisition, the second crystal 
rotates continuously at a roughly constant speed. Fluctuations in the step size however, due to backlash in the gears and non-uniformity in the stepping motor magnetic field, lead to small variation of the time spent in each bin. 
The scanning range is divided into a number of bins of identical size (typically 100 bins of 5.7"). The counts are stored in a bin when the angle value measured by the encoder is contained between the  minimum and maximum angle defined for the bin. The content of each bin is divided by the time spent in the bin to insure proper normalization. The time during which the first axis wanders too far away from the set position, leading to a stop in x~ray collection, is measured and subtracted from the acquisition time for a given bin.
In a typical spectrum the time spent in a bin is around \unit{13.5}{s} and can fluctuate between  \unit{10}{s} and  \unit{17}{s}. A typical spectrum is recorded in \unit{10}{min} in the parallel mode and in \unit{20}{min}  in the dispersive mode.

A Xe (90 \%) and methane (10 \%) gas filled proportional counter detector is mounted on a Newport RV 240PP rotation stage with an axis of rotation concentric with the second crystal vertical axis. The detector has a \unit{50}{\mu m} thick Be window and has an active area of about \unit{12\times 25}{mm^2}. The detector is operated at a high voltage of about \unit{2000}{V} with and external power supply. The detector signal is processed by an ORTEC 142PC low noise charge-sensitive preamplifier and an ORTEC 572 spectroscopy amplifier with a shaping time of \unit{6}{\mu s}. An ORTEC window and scalar module is used to generate TTL pulses when the signal amplitude corresponds to the expected x-ray energy. These pulses are accumulated by a 6602 PC card from National Instrument. A Labview program pilots the microstepping motor control unit for both  axes and   detector rotation stages. The same program reads the AWE 1024 controller trough a GPIB bus and the first crystal encoder with an Heidenhain IK 220 PC card. The program uses the reading from the first axis encoder to maintain the angle, while scanning the second axis angle and acquiring the counts from the 6602 card. The program displays and updates a plot of the crystals angular positions and of the spectrum recorded during a complete measurement cycle.
%

\begin{figure}
[htb]
\centering
\includegraphics[width=\columnwidth]{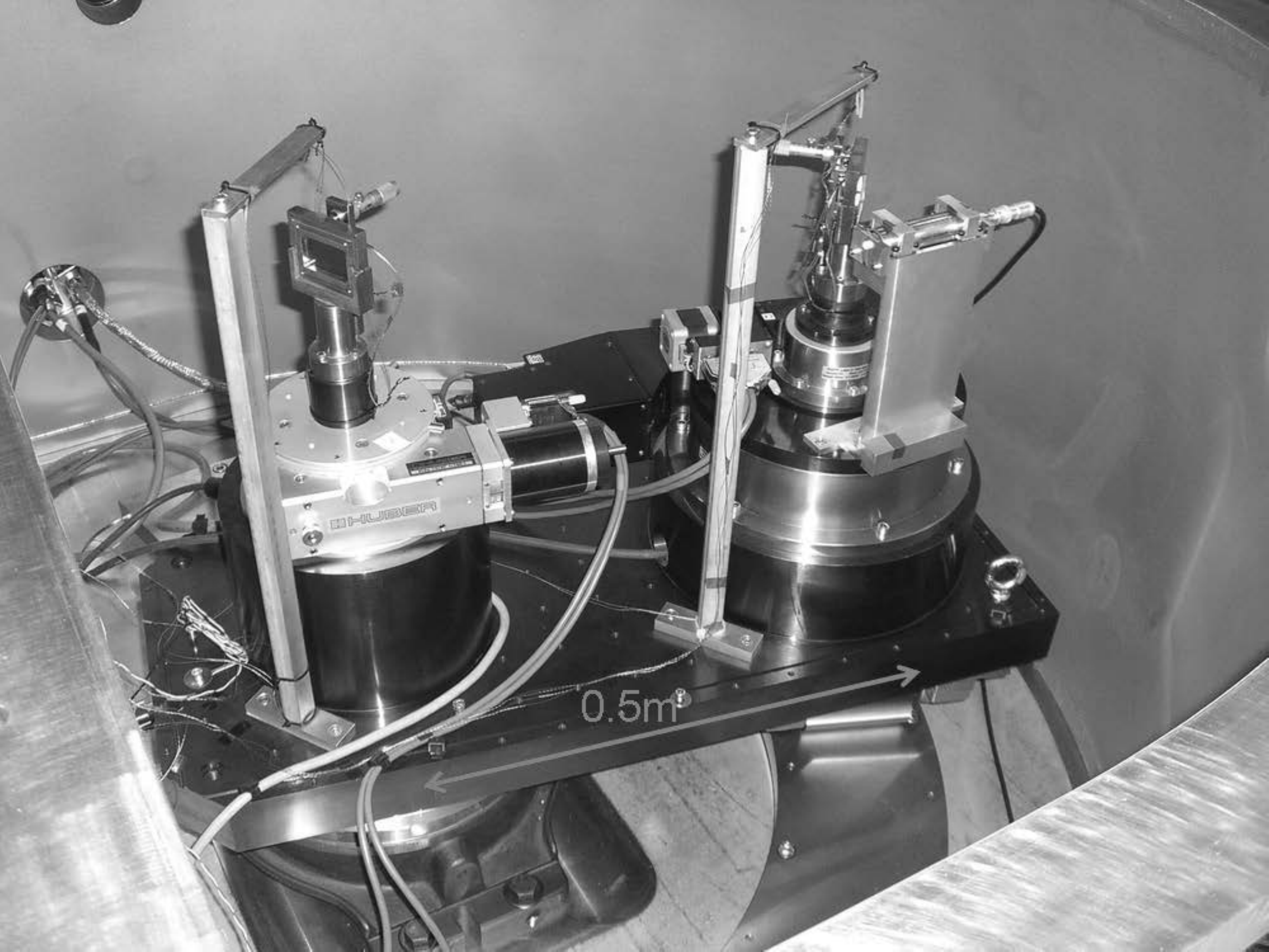}
\caption{General view of the spectrometer.}
\label{fig:spectro-photo}
\end{figure}

\begin{figure*}
[htb]
\centering
\begin{tabular}{c}
\includegraphics[width=\textwidth]{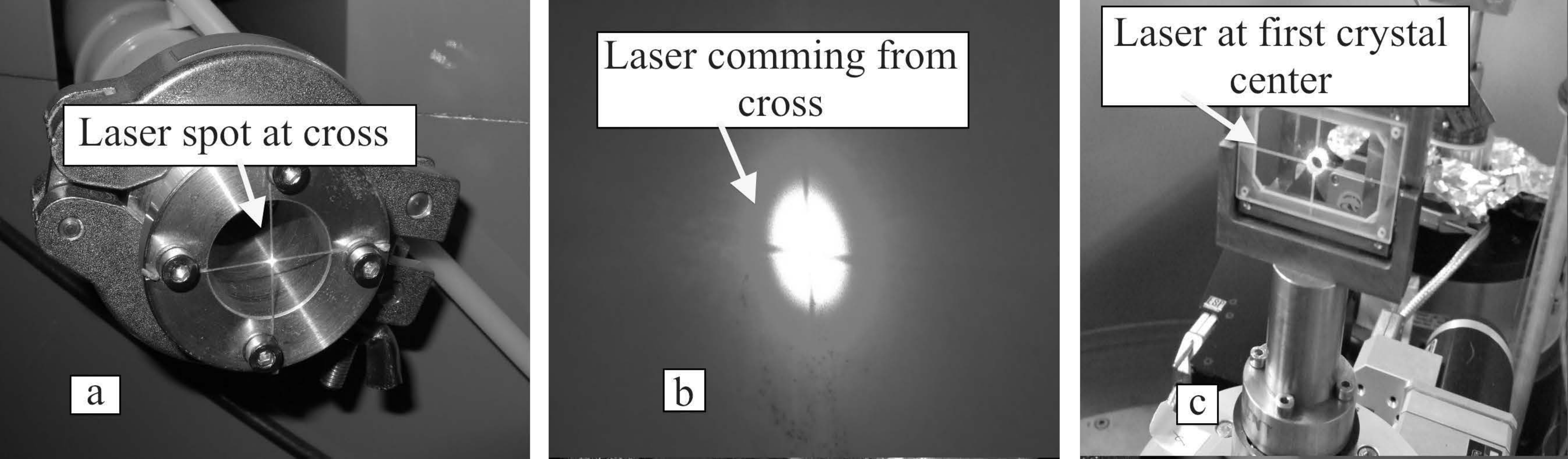} \\
\includegraphics[width=\textwidth]{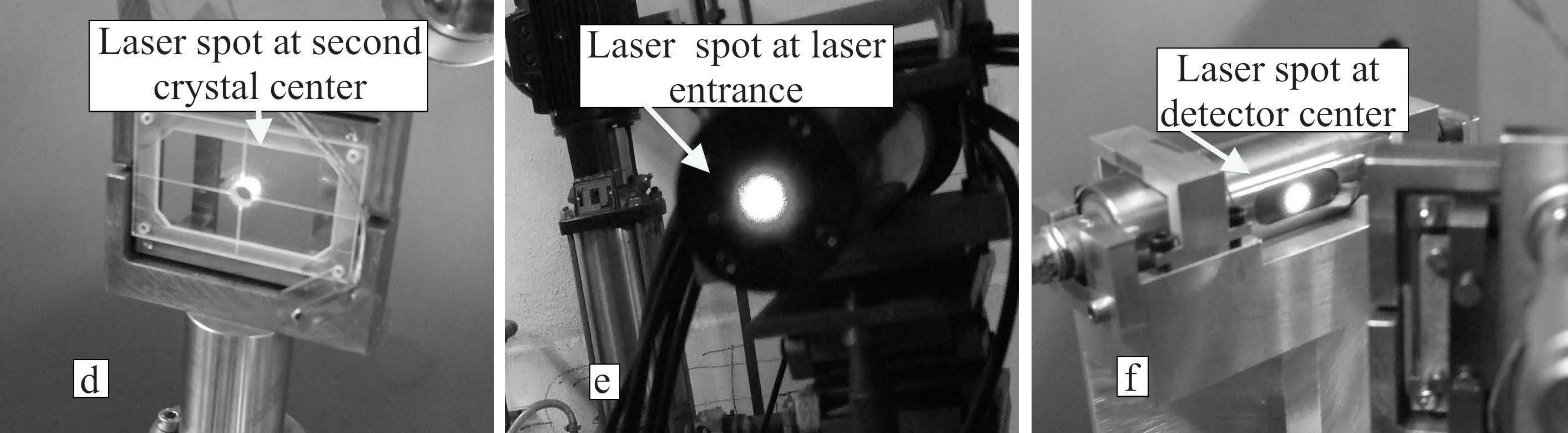} 

\end{tabular}
\caption{The laser beam is aligned with: a) cross located at back side of the ECRIS source; b) the cross located at front side of the ECRIS source (b is a picture of the out-going laser on a screen); c) the center of the first crystal; and d) the center of the second crystal. Then, a high-quality mirror is inserted  in place of the second crystal. The encoder offset is set by making the laser beam go back to the starting point e).  After moving the table, the second crystal is set in the measurement position and the detector is aligned with the laser f).}
\label{fig:crystal-aliglaser}
\end{figure*}

The temperature of the crystals is measured to \unit{0.1}{°C} accuracy using a calibrated Pt100 thermistor. This thermistor is also used to regulate the temperature of the crystal.  A heating element is pressed between two thin copper plates, which are applied to the back of the crystal (Fig. \ref{fig:crystal-support}). A \unit{100}{\mu m}-thick soft graphite foil assures a good thermal contact between the crystal and the Cu plate in vacuum. Water cooling  is applied to the rotary stages stepping motors, in order to provide sufficient heat loss when the spectrometer is under vacuum. A feedback loop controls the power in the heating element using a proportional-integral-derivative (PID) controller. The maximum allowed fluctuation in the course of one measurement is \unit{0.2}{°C}. The temperature of both crystals is also recorded during the scans with each data point registered in the data files.

\subsection{Alignment procedure}
\label{subsec:align}

\begin{figure*}
[htb]
\centering
\includegraphics[width=\textwidth,trim =1.cm 9.cm 1.cm 6.cm,clip]{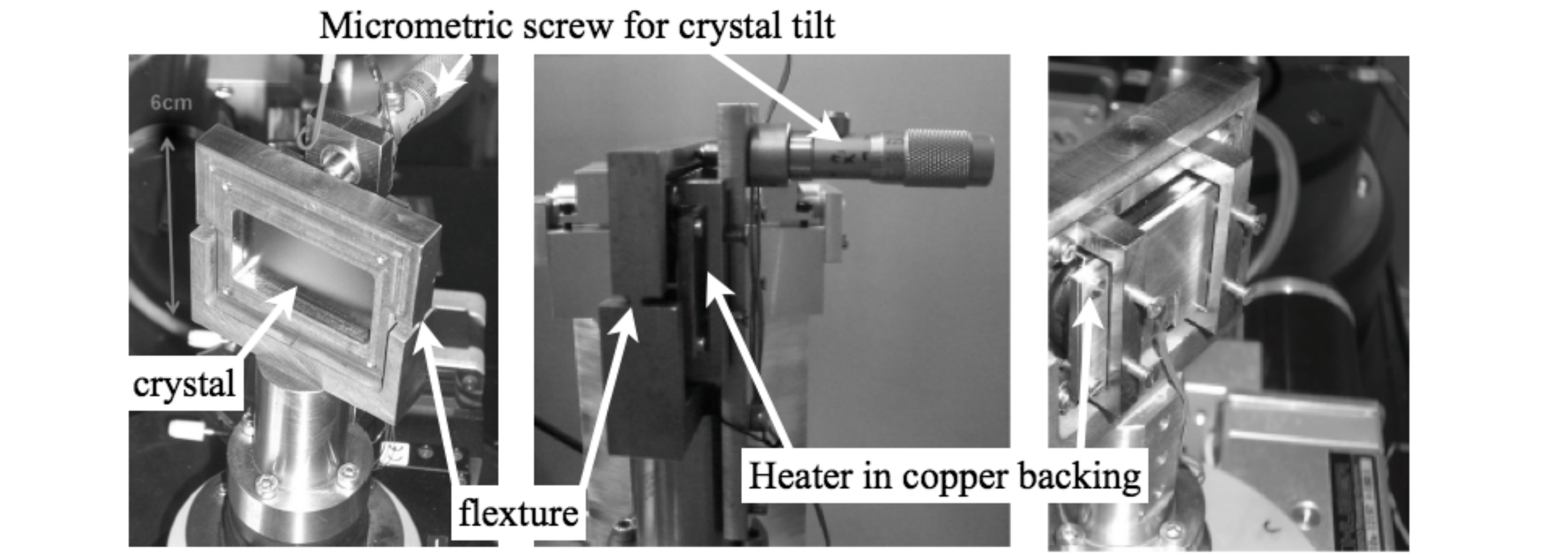}
\caption{Details of the crystal support, with the vertical tilting system (composed of a flexture and of a micrometric screw) and the heater plate for temperature control. The copper plate ensures uniform repartition of heat on the back of the crystal. A graphite foil is positioned between the crystal and the copper plates to improve thermal contact under vacuum.}
\label{fig:crystal-support}
\end{figure*}

The DCS must be carefully aligned  with respect to the SIMPA axis to optimize the spectrometer throughput and to allow finding easily the lines that are to be measured. The quality of the vertical alignment is  very important for reducing systematic errors. 
The procedure is the following. First, two carefully machined cylindrical pieces with crosshairs are placed on the flanges at the exit of the source in place of the Be window (see Fig. \ref{fig:simpa-photo-connexion}) and at the end of the beam line on the other side of the source (on the alignment port of the \unit{1.5}{T} dipole magnet Fig.~\ref{fig:crystal-aliglaser}, a). Both ports have been aligned with the source before. A theodolite, equipped with angular encoders of arcsecond accuracy  and  with an electronic tiltmeter is then positionned so that it is on a straight line with respect to the crosshairs. The horizontality of the axis can be verified to a few seconds of arc  using the tilt-meter. The spectrometer chamber is then equipped with crosshairs on the entrance and exit flange. The theodolite is then used to align horizontally and vertically the chamber.  A lead diaphragm, slightly smaller than the detector entrance window is installed on the entrance port of the spectrometer chamber, and its alignment checked. The spectrometer table is then rotated so that both crystal supports are aligned with the source axis. An alignment laser is then set to go through the crosshairs (see Fig.~\ref{fig:crystal-aliglaser}, a to d). A high-quality mirror is  installed in place of the second crystal. The axis is then rotated until the laser is reflected back onto itself. The verticality of the mirror is adjusted using a micrometric screw (Fig. \ref{fig:crystal-support}). The crystal support rotates on an axis going through the front of the mirror, using a system of flexure hinges \cite{paw1965}.  This enables the substitution of the mirror by a crystal without losing the vertical alignment. The accuracy of this alignment is defined with the precision with which the laser beam can be centered when reflected back onto itself (see Fig.~\ref{fig:crystal-aliglaser}, e). This is around \unit{2}{mm} over a distance of \unit{16}{m}, i.e., 13''.
The angle on the encoder of the second crystal axis is then set to \unit{90}{degrees} to provide a logical reference angle for the measurements.

Once the second crystal support is aligned, the same procedure is repeated on the first crystal support. At this point both mirrors are parallel to each other and perpendicular to the source axis. This is called the nominal alignment position. The first crystal is then rotated to the Bragg angle value corresponding to the transition to be measured, and the spectrometer table will be rotated until the laser beam hits the center of the second mirror and is reflected back onto itself. The second crystal is then rotated to the Bragg angle so that both mirrors are parallel. The x-ray detector is then positioned on the laser beam to mark the detector position for the parallel or non-dispersive mode (see Fig.~\ref{fig:crystal-aliglaser}, f). Finally the second crystal is rotated so that it is at the correct Bragg angle for the dispersive mode and the detector is moved to the correct position to check if it is correctly centered on the laser beam and to mark its position in the non-dispersive mode.

The horizontality of the various components was checked with a Wyler Clino 2000 tiltmeter to a precision of a few seconds of arc, and the verticality of the crystals with a Wyler Zerotronic sensor to the same accuracy.

The mirrors are then replaced by the crystals, and the spectrometer is set up for x-ray measurements. The ECRIS is started, and an x-ray picture is taken in front of the first crystal to check that it is uniformly illuminated. The total uncertainty associated with the alignment procedure is \unit{0.01}{degrees}.
Procedures to check this alignment are presented in Secs. \ref{sec:uncertain} and \ref{sec:results}.


\subsection{Crystals preparation and measurement}
\label{subsec:crystals}
In order to obtain an absolute energy measurement with the DCS, it is necessary to know the crystal lattice spacing with high accuracy. Four silicon crystals have been manufactured at NIST for the Paris DCS, two with Miller indices (111)  and two with (220). Polishing procedures that lead to optical-quality surfaces (e.g., diamond powder polishing) damage the crystal surface and are not satisfactory for obtaining high-quality crystals  for x-ray spectroscopy. Chemo-mechanical polishing (CMP) has been shown to lead to  a somewhat broader distribution of lattice spacing values \cite{dkob1999}. The crystals were attached to a support with wax, oriented using a crystal x-ray spectrometer 
and lapped, using SYTON, a colloidal silica slurry.  The lapping was performed so that the cut angle (angle between crystal planes and crystal surface) is smaller than \unit{10}{arcseconds} to reduce asymmetric cut corrections to a negligible value \cite{jam1948}. The crystals were then etched to remove strains and surface damages and minimize lattice spacing dispersion. The crystals have thus a slightly frosted aspect, making the surface rather diffusive for laser light.

\subsubsection{Description of the measurement}
\label{subsubsec:crystal-prep}

All four crystals were cut from a boule obtained from Wacker-Siltronic. A small test crystal was prepared from the same boule for measurement of the lattice spacing, using the so-called ``delta-$d\,$'' spectrometer from NIST.  \cite{khdn1994}  The physical separation between the  ``delta-$d\,$'' diffraction crystal and the DCS diffraction crystals was kept as small a possible so that any variation in lattice spacing along the boule will have negligible influence on the determination of the lattice spacing of the DCS diffraction crystals. 
 Although it is expected that the lattice spacings of the two samples are identical, a relative uncertainty component of $10^{-8}$ is included in the lattice spacing uncertainty to account for possible sample-to-sample variations \cite{kohd1999}. The ``delta-$d\,$'' crystal was cut from a \unit{18}{mm}$\times$\unit{12}{mm}$\times$\unit{6}{mm} sample with a thin lamella for diffraction as the top half and a base for mounting as the bottom half. The ``delta-$d\,$'' crystal was etched to a lamella thickness near \unit{0.450}{mm}, which was determined by fitting the  ``delta-$d\,$''  machine spectra with theoretical diffraction profiles.
 
The test crystal was attached to a PZT tipper using soft wax and the crystal/PZT tipper assembly was mounted on the ``delta-$d\,$'' translation sled. An other sample crystal, the reference crystal, cut from the WASO04 silicon boule, was also mounted on the translation sled. The WASO04 boule, also grown by Wacker-Siltronic, is a nearly perfect single crystal natural silicon material that was specifically grown for the International Avogadro Project \cite{khdn1994,kohd1999,fwkm2005,mtn2008,aabb2011,mmfk2011}. The lattice spacing of the WASO04/NIST reference crystal was determined as part of the International Avogadro Project \cite{mmfk2011}.


The comparison measurements were recorded in the period from January 9 to 23, 2006. The relative difference in lattice spacing between the DCS test crystal and the WASO04/NIST reference crystal was measured as well as the variation in lattice spacing over the central  \unit{6} of the DCS crystal.

\subsubsection{Result of lattice comparison measurement}
\label{subsubsec:lattice-comp}
The ``delta-$d\,$''  spectrometer measures the small differences in Bragg angle between two crystals, from which the lattice spacing difference of the two crystals is inferred. Silver K$\alpha$ radiation is diffracted in a two-crystal transmission non-dispersive geometry and the recorded profiles are fit with theoretical dynamical diffraction profiles. A complete description of the spectrometer and the measurement procedures is available in Ref.  \cite{khdn1994}.

Profiles were recorded with the second crystal position alternately being occupied by the test and reference crystal. The first crystal was rotated both clockwise and counterclockwise. Temperatures of the first crystal and of the test and reference crystals were measured at each data point and small corrections for temperature are made to the raw data before fitting. Over a  \unit{24}{hour} hour period, typical temperature changes of about  \unit{0.010}{°C} were noted.

The data used to obtain the lattice spacing difference between the two crystals included 150 data scans recorded over 10 days.  The measured lattice spacing difference is  $(d_{\mathrm{test}}-d_{\mathrm{Ref.}})/d_{\mathrm{Ref.}} =(-2.3 \pm 0.5) \times 10^{-8}$ where the uncertainty includes a statistical component 
($2\times 10^{-9}$) and systematic components associated with crystal temperature measurements ($3 \times 10^{-9}$), crystal alignment ($10^{-9}$), and location of x-ray paths and crystals ($3 \times 10^{-9}$). 

In order to take into account variations along the Wacker-Siltronic boule, we include a relative uncertainty component of  $10^{-8}$ and convert the measurement reported in the previous paragraph to the final result for the DCS spectrometer crystals  $(d-d_{\mathrm{WASO04/NIST}})/d_{\mathrm{WASO04/NIST}} = (-2.3 \pm 1.1) \times 10^{-8}$. 

The variation of the lattice spacing along the surface of the DCS test crystal was measured by comparing the central region with the \unit{\pm3}{mm} regions surrounding it. The relative change in lattice spacing along this  \unit{6}{mm} region was measured to be $8.6 \times 10^{-9}$. This variation is consistent with the $10^{-8}$ relative uncertainty component that has been attributed to the lattice parameter variation along the Wacker-Siltronic boule.

\subsubsection{Absolute lattice parameter value}
\label{subsubsec:lattice-val}
There have been several new measurements of the  $d_{220}$ lattice spacing of natural silicon in the past five years \cite{fmm2008,mmk2009,mmkf2009}. In addition, the lattice parameter of an ideal single crystal of naturally occurring Si, free of impurities and imperfections is one of the quantities that is determined in the CODATA recommended values of the fundamental physical constants. The variation of the $d_{220}$ value between the 2006 \cite{mtn2008} and 2010 \cite{CODATA2010} CODATA recommended values is more than 3 times the stated uncertainty. In order to compare lattice parameter values of different crystals, corrections for measured C, O, and N impurity concentrations are taken into account. From this collection of lattice parameter values, a straightforward approach to an absolute lattice spacing value for the WASO04/NIST reference crystal is not obvious. 


Fortunately, in a 2011 publication a $d_{220}$ value for the specific WASO04/NIST reference crystal was determined \cite{mmfk2011}. The lattice parameter of a sample taken from a specific location (\unit{87}{cm} from crystal seed) in the WASO04 boule was carefully measured. Then corrections were made for the variation in the C, O, and N impurity concentrations along the WASO04 boule to determine the lattice spacing of the WASO04/NIST reference crystal (\unit{143}{cm} from the crystal seed). The value of the lattice spacing of the WASO04/NIST reference crystal is \unit{d_{220} = 192.0143374(10) \times 10^{-12}}{m} in vacuum at \unit{20}{°C}. The relative uncertainty of this value is $0.5 \times 10^{-8}$. This value can be adjusted to the laboratory temperature by using the expansion expression for natural silicon
\begin{equation} 
\label{eq:si-dilat}
\frac{\Delta d}{d} =  \alpha_0(T - 20) + \alpha_1(T - 20)^2,
\end{equation}
 where $T$ is the laboratory temperature in °C, \unit{\alpha_0 = 2.5554 \times  10^{-6}}{°C^{-1}}, and \unit{\alpha_1 =  4.58 \times 10^{-9}}{°C^{-2}} (Ref.  \cite{sab2001}) 
 at \unit{22.5}{°C} in vacuum the WASO04/NIST reference crystal lattice spacing is  \unit{d_{220}  = 192.0155696(10) × 10^{-12}}{m}. 

%

Finally, the measured lattice spacing difference between WASO04/NIST reference crystal and the DCS crystal material is used to calculate the absolute lattice spacing of the DCS crystals \unit{d_{220} = 192.0155651(23) \times 10^{-12}}{m} and \unit{d_{111} = 313.5601048(38)  \times 10^{-12}}{m} in vacuum at \unit{22.5}{°C}. The results and uncertainties for the Si (220) crystals are summarized in Table \ref{tab:si-2d-final}. When these crystals are used in a laboratory environment for diffraction measurements, the above lattice parameter values should be adjusted for the temperature of the crystals and the laboratory air pressure. The expansion correction is given in Eq. \eqref{eq:si-dilat} and the compressibility correction is
\begin{equation}
\label{eq:si-press}
 \frac{\delta d}{d} = -\epsilon p\, ,
 \end{equation}
 where $p$ is the laboratory pressure in Pascals, \unit{\epsilon = 0.3452 \times 10^{-6}}{atmosphere^{-1}} \cite{mcs1953,nye1957}. For a pressure of \unit{1}{atmosphere}, the relative correction is approximately $-3.4 \times 10^{-7}$.


\begingroup
\begin{table*}[tb]
\begin{center}
\caption{\label{tab:si-2d-final}
Lattice spacing value for the Si (220) crystals of the spectrometer at \unit{22.5}{°C} in vacuum. The Si (111) values can be deduced by multiplying with the factor $\sqrt{8/3}$. Numbers in parenthesis are uncertainties.
}
\begin{tabular}{ldd}
		&	\multicolumn{1}{c}{Value (\AA)}		&	\multicolumn{1}{c}{Relative accuracy (ppm)}	\\
\hline							
WASO04/NIST reference crystal (\unit{22.5}{°C}, vacuum)	&	1.920155696(10)	&	0.005	\\
``delta-$d\,$'' measurement	&	-0.000000045(21)	&	0.011	\\
DCS Si (220) crystals (\unit{22.5}{°C}, vacuum)	&	1.920155651(23)	&	0.012	\\
\end{tabular}
\end{center}
\end{table*}
\endgroup


%
%

\section{Simulation of  the DCS}
\label{sec:theoryDCS}
We have developed a ray-tracing program to obtain theoretical line profiles for the DCS, in the dispersive and non-dispersive modes. The results of this simulation program are used to analyze the experimental data. The program is based on the Monte-Carlo method and includes all relevant geometrical components of the experiment, as shown on Fig. \ref{fig:simpa-DCS-geom},  along with the crystal reflectivity curve  calculated by dynamical diffraction theory (see, e.g., Ref.  \cite{zac1967}) using XOP \cite{sad1998,sad2004,sad2004a} and checked with  X0h \cite{stepanov,las1991}. This makes the simulation code capable of taking into account multiple reflections in the crystal and  corrections to the Bragg law, such as the index of refraction corrections and energy-dependent absorption. A  distribution function is assigned to each x-ray line included in the simulation, to take into account  its natural width (Lorentzian functions) or Doppler broadening (Gaussian function) or both (Voigt function). The simulation is thus capable of providing a line-width analysis for our experimental spectra.

\begin{figure}
[htb]
\centering
\includegraphics[clip=true,width=8cm]{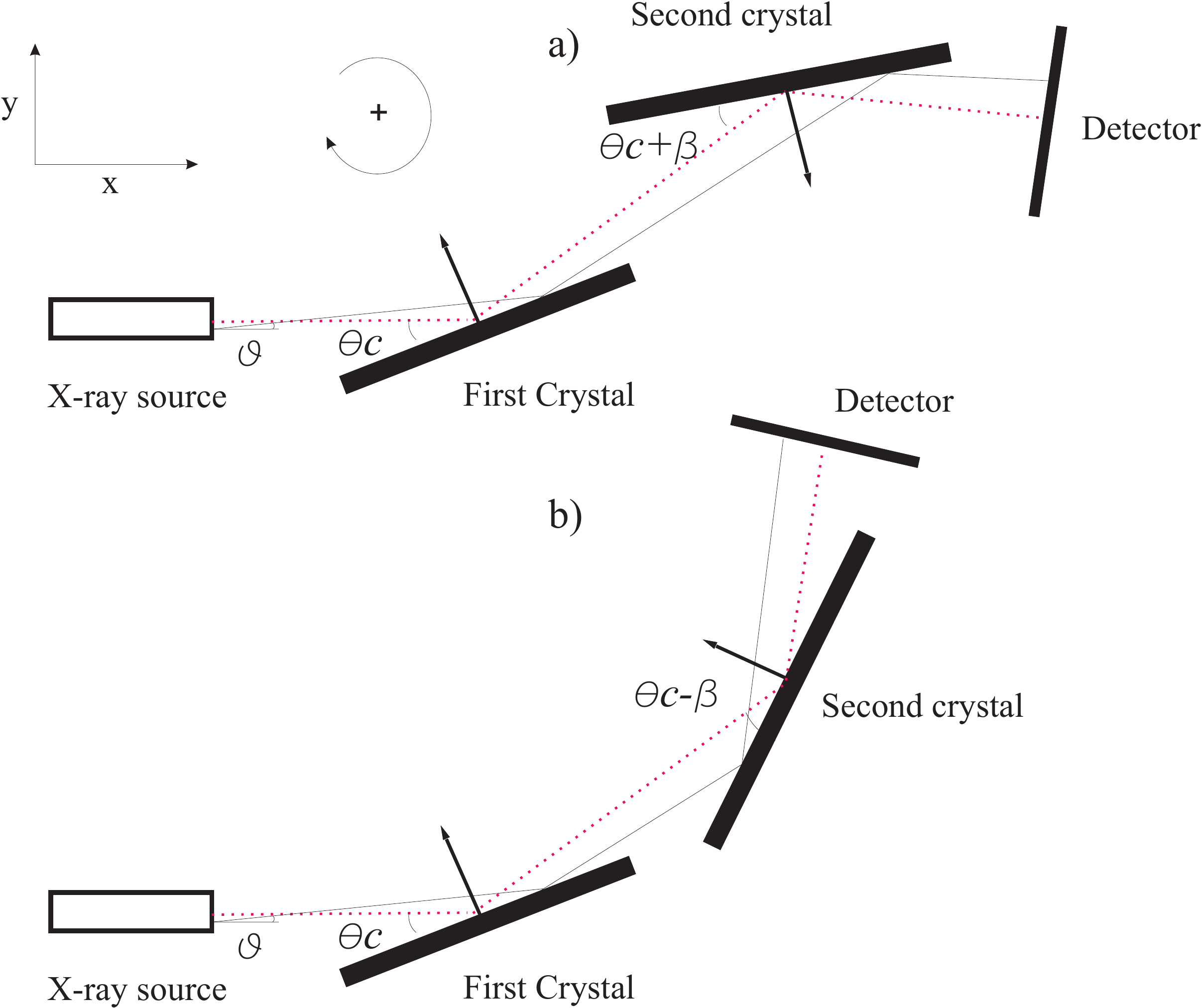}
\caption{Geometry of the DCS in the horizontal plane. a) and b) refers to the non-dispersive and the dispersive positions respectively. The  dotted line defines the central
beam named ``central line'' in the simulation model. $\theta$ is the horizontal deviation of the x rays compared to the central line, $\theta_{\mathrm{C}}$ is the central line's angle with respect to the first crystal and $\theta_{\mathrm{C}}\pm \beta$ is the central line's angle with respect to the second crystal in the non-dispersive and dispersive modes respectively.  The crystallographic planes of the crystals are defined by their normal vectors.}
\label{fig:geo_DCS_Hor}
\end{figure}

A simulated line profile is represented by the number of rays hitting  the detector as a function of the second crystal angle. This curve is sometimes called the rocking curve.  \cite{all1932}
The non-dispersive profile, represented by ($n_{1}$, $-n_{2}$) in Allisson's notation \cite{aaw1930}, where $n_{i}$ is the
order of diffraction on the $i$th crystal, is obtained by scanning the second crystal in the case in which both crystals are parallel, as shown in Fig.~\ref{fig:geo_DCS_Hor}, a). This profile is called non-dispersive since
each bin in the rocking curve has contributions from all wavelengths accepted by the first crystal and reaching the second crystal. The peak in this profiles indicates that the crystallographic planes of both crystals are parallel.

The dispersive profile noted ($n_{1}$, $+n_{2}$) which corresponds to the geometry represented in Fig.~\ref{fig:geo_DCS_Hor}, b)  provides a peak for the case of a (quasi) monochromatic incoming x-ray line. The peak profile in this case is a convolution product of the instrument response function and the natural line shape. The observed intensity in this configuration is much lower than in the  ($n_{1}$, $-n_{2}$) configuration, as each angle corresponds to only one wavelength, within the width of the crystals reflectivity curve. Up to now we have only performed measurements in first order, so we will restrict our analysis to the ($1$, $-1$) and ($1$, $+1$) cases.

\begin{figure}
[htb]
\centering
\includegraphics[clip=true,width=8cm]{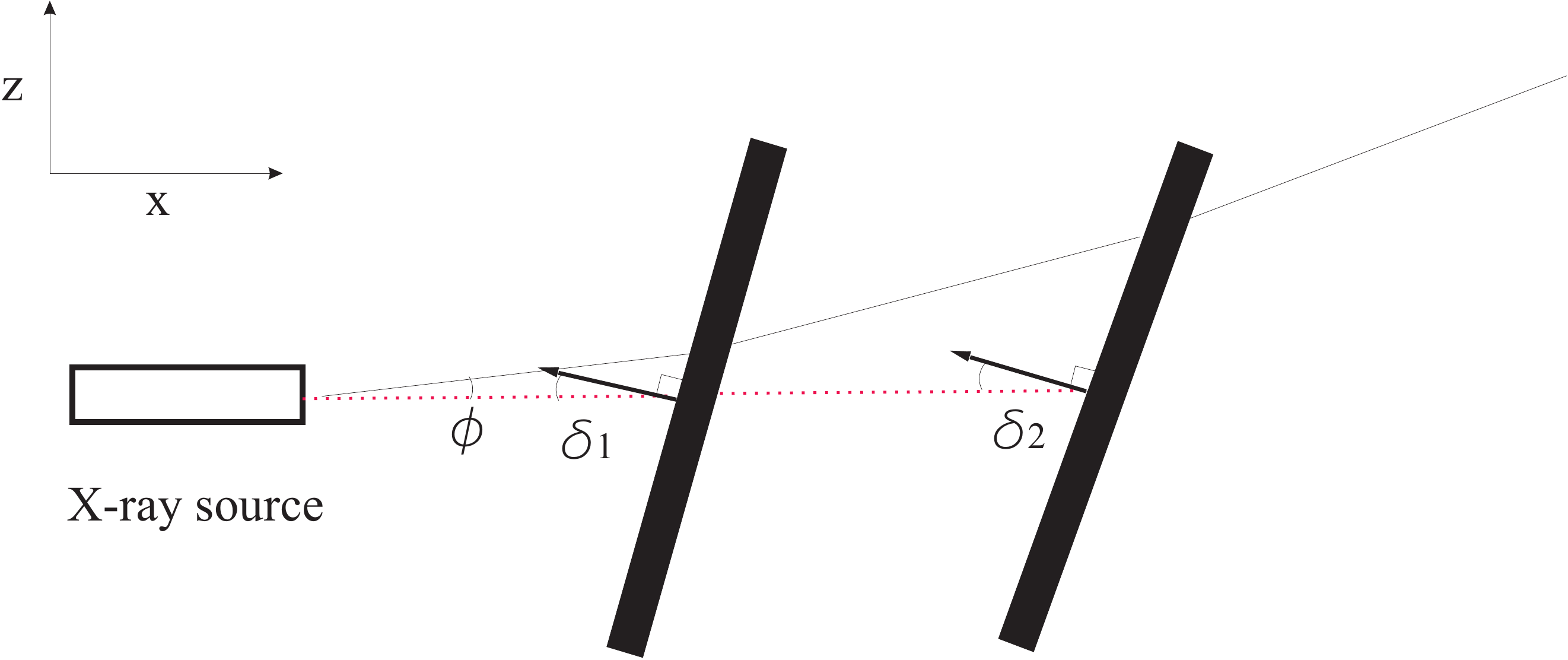}
\caption{Geometry of the DCS in the vertical plane in the nominal alignment position (see section II. C) when the spectrometer table is placed to be parallel to the axis of the source and the crystals are rotated to be perpendicular to this same axis. This is not an actual measurement position, but  serves as an example for the crystals tilts and the beam. The dotted line defines the central
beam (or central line) used in the simulation model. $\phi$ is the vertical divergence angle of the x-ray beam at the source, $\delta_1$ and $\delta_2$ are the vertical tilt angles of both crystals respectively. (The tilts and vertical divergence angles are exaggerated on this figure.)}
\label{fig:geo_DCS_ver}
\end{figure}

\begin{figure}
[htb]
\centering
\includegraphics[clip=true,width=8cm]{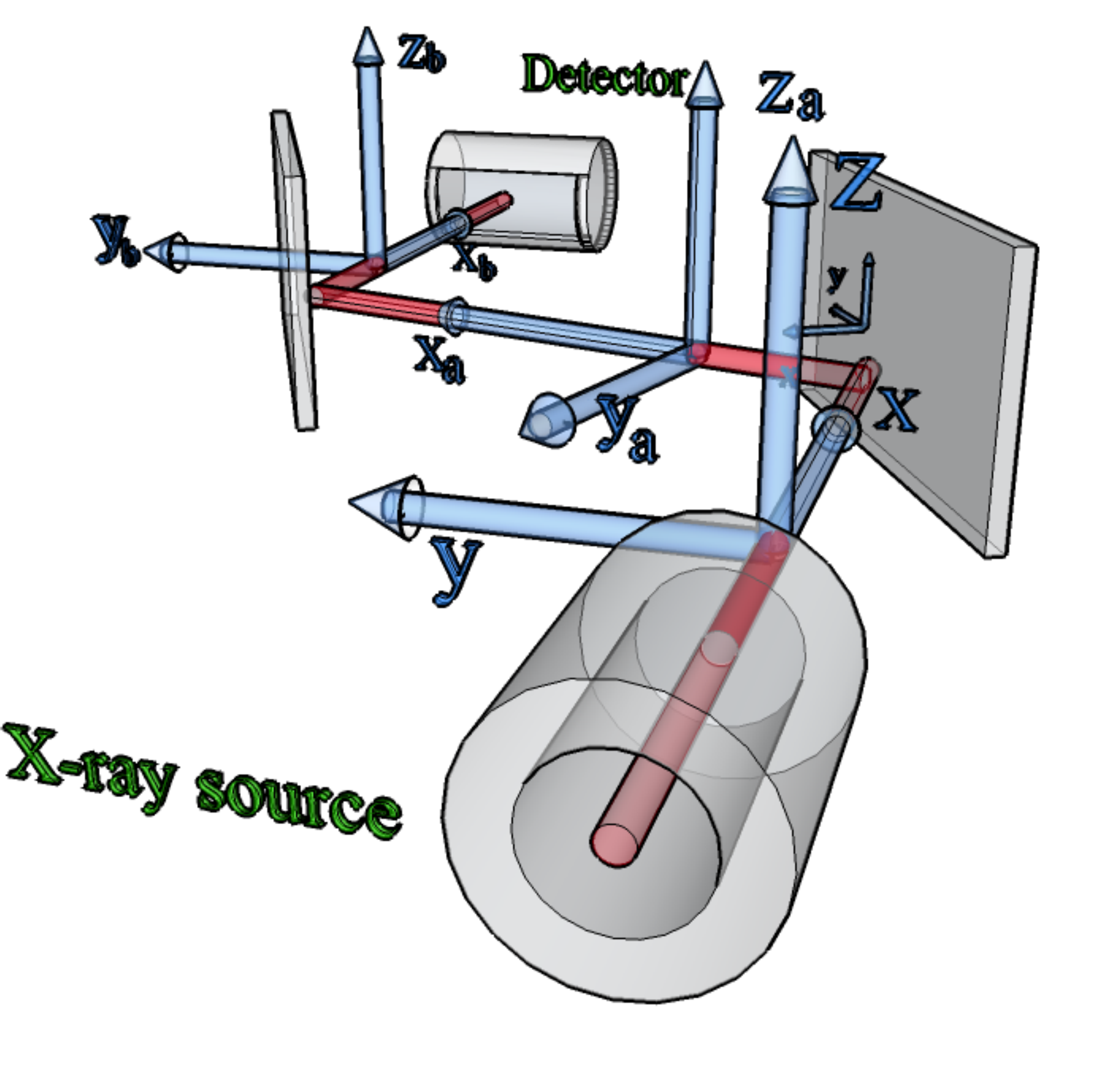}
\caption{(Color online) Geometry of the DCS in a non-dispersive setup with the central line and the tri-orthogonal $xyz$ axis along the central line.}
\label{fig:axes_def}
\end{figure}

The vertical geometry of the DCS in the nominal alignment position is shown in  Fig. \ref{fig:geo_DCS_ver} to demonstrate the vertical divergence angle $\phi$ and the crystal tilt angles $\delta_1$ and $\delta_2$ used in the simulation. A succession of three $xyz$ (orthogonal) coordinate systems are defined that follow the central line in the simulation (see Fig. \ref{fig:axes_def}). Each randomly generated ray will be represented in these coordinate systems within the three different parts of the experiment. The three coordinate systems are shown for the non-dispersive case in Fig. \ref{fig:axes_def}. 
The central line, that is the line connecting the geometrical centers of the different components of an ideally aligned spectrometer defines the $x, x_a, x_b$ axes of the three successive coordinate systems respectively.

%

%
%

%

Misalignments of successive components of the experiment defined in Figs. \ref{fig:geo_DCS_Hor} and \ref{fig:geo_DCS_ver} are taken into account in the simulation with respect to these three ideal $xyz$ coordinate systems shown on Fig. \ref{fig:axes_def}.
We define the angle  $\theta_{\mathrm{T}}$ as the horizontal angle between the ion source axis and the plane defined by the two crystal axes. When the experiment is properly set, we should have $\theta_{\mathrm{T}}\approx2\theta_{\mathrm{C}}$ and $\theta_{\mathrm{C}}\approx \theta_{\mathrm{B}}$, the Bragg angle (see Fig. \ref{fig:geo_DCS_Hor} for the other definitions).


A simulated rocking curve is calculated using $\approx 10^6$ rays, each defined by generating a set of three $xyz$ coordinates and two angles $\phi$ and $\theta$ with a uniform random generator, for successive values of the scanning angles $\beta$. A simulated spectrum is created by counting the number of x rays reaching the detector for a given value of $\beta$.
In order to save computer time the values of $\phi$ and $\theta$ are constrained in the range (U$ \left[ \theta_{\mathrm{min}},\theta_{\mathrm{max}} \right] $, U$\left[ \phi_{\mathrm{min}},\phi_{\mathrm{max}} \right] $), where the  angles  $\theta_{\mathrm{min}}$, $\theta_{\mathrm{max}}$, $\phi_{\mathrm{min}}$ and $\phi_{\mathrm{max}}$ 
are given by the successive collimators between the source and the first crystal (see Fig. \ref{fig:simpa-DCS-geom}).


The ray direction is expressed by the cartesian components of the unitary vector $\bm{\hat{e}}$, 

%

\begin{eqnarray}
\hat{e}_x &=& \cos(\phi) \cos(\theta)~ , \nonumber  \\
\hat{e}_y   &=& \cos(\phi) \sin(\theta) ~ ,    \nonumber \\
\hat{e}_z  &=& \sin(\phi) \qquad ~ ~ ~ . 
 \label{eq:vect_dire}
\end{eqnarray}
Furthermore, the initial position $yz$ at the sourceexit is defined by a fixed
uniform random distribution (U$\left[ -R_c, R_c \right] $, U$\left[
  -R_c, R_c \right] $), where $R_c$ is the source tube radius. If a
position is generated outside the region $y^2+z^2<R_c^2$, it is
discarded and another point is generated. This procedure was used for saving computer time since evaluation of trigonometric functions is minimized.




The  position of the ray at the crystal in the plane $y'z'$, perpendicular to the source axis, which includes also the first crystal axis of rotation, is given by 

\begin{eqnarray}
y'&=&y+L \tan(\theta)~ , \nonumber  \\
z'&=&z+L \tan(\phi)~ , 
\label{eq:trans}
\end{eqnarray}

where $L$ is 
the distance between the source and the first crystal. The position $y''z''$ on the surface of the first crystal is given by the projection of the position $y'z'$ over the surface axes

\begin{eqnarray}
y''&=&y'\frac{\cos(\theta)}{\cos(\theta+\frac{\pi}{2}-\theta_{\mathrm{C}})} ~ , \nonumber  \\
z''&=&z'\frac{\cos(\phi)}{\cos(\phi+\delta_1)} ~ .
\label{eq:proj_sur}
\end{eqnarray}

The angle between the ray and the crystallographic plane of the first crystal is given by

\begin{equation}
\alpha_1= \arcsin(-\bm{\hat{e}}  \cdot \bm{\hat{n}}_1)  ~ ,
\label{eq:alph}
\end{equation}

where $\bm{\hat{e}} $ is the ray vector direction (Eq. \eqref{eq:vect_dire}) and $\bm{\hat{n}}_1$ is a unitary vector perpendicular to the crystallographic planes of the first crystal expressed by
\begin{eqnarray}
\hat{n}_{1x} &=& -\cos(\delta_1) \cos(\theta_{\mathrm{C}}) ~ , \nonumber  \\
\hat{n}_{1y }  &=&~ ~ \cos(\delta_1) \sin(\theta_{\mathrm{C}}) ~ ~ ,    \\
\hat{ n}_{1z}  &=& ~ ~ \sin(\delta_1) \qquad \qquad  . \nonumber
\label{eq:vect_dire_1}
\end{eqnarray}
Therefore, the direction of the reflected ray is given by

\begin{equation}
\bm{\hat{e}'} =\bm{\hat{e}}-2 (\bm{\hat{e}} \cdot \bm{\hat{n}}_1)  \bm{\hat{n}}_1~ .
\label{eq:r2}
\end{equation}

If the ray position is within the boundaries of the crystal, a wavelength  $\lambda$ is generated using a Lorentzian random number generator. The normalized Lorentz function is given by
\begin{equation}
L(\lambda, \lambda_1, \Gamma_1)=\frac{\frac{\Gamma_1}{2\pi}}{\left(\lambda-\lambda_1\right)^2+\left(\frac{\Gamma_1}{2}\right)^2}~,
\label{eq:lore}
\end{equation}
where $\lambda_1$ is the transition wavelength and  $\Gamma_1$ is the natural line width  (FWHM) associated with the decay lifetime. The method used for generating the
 random number  with a Lorentzian distribution is the inverse method \cite{pftv1986}. The effect of the Doppler broadening is obtained by  generating a wavelength $\lambda'$ with a Gaussian random
number generator centered at the wavelength $\lambda$
\begin{equation}
 G(\lambda', \lambda, w)= \frac{2}{w}\sqrt{\frac{\ln(2)}{\pi}} \exp\left(-\frac{\left(\lambda'-\lambda\right)^2}{\left( \frac{w}{2\sqrt{\ln(2)}}\right)^2}\right),
 \label{eq:gauss}
\end{equation}
where the  FWHM  $w$ is given by the velocity distribution of the ions. The Gaussian random number generator is implemented using the Box-Muller method, also based on the inverse method \cite{pftv1986}. In that way we generate a wavelength  
$\lambda'$ corresponding to a Voigt profile (the convolution of the Lorentz and Gaussian distributions).

The Bragg angle, $\theta _{\mathrm{B}}$, is related to the wavelength $\lambda'$ by the well-know relation 
%
\begin{equation}
\lambda'=2d\sin \left( \theta _{\mathrm{B}}\right)  ,
\label{eq:lamd}
\end{equation}%
where $d$ is the lattice spacing. 

The  temperature dependence of the lattice spacing is given by 
\begin{equation}
d(T)=d_{22.5}(1+(T-22.5)\alpha(T))\, ,
\end{equation}
where the temperature is given in degrees Celcius, $\alpha(T)$ is the dilatation coefficient and $d_{22.5}$ is the lattice spacing at \unit{T=22.5}{°C}. Since we are dealing with small temperature changes, we neglect the temperature dependence of $\alpha(T)$. The quantities $\alpha\left(T\right)$  and $d_{22.5}$  are given in Subsec. \ref{subsubsec:lattice-val}.
%

%

\begin{figure}[tb]
\centering
\includegraphics[clip=true,width=\columnwidth]{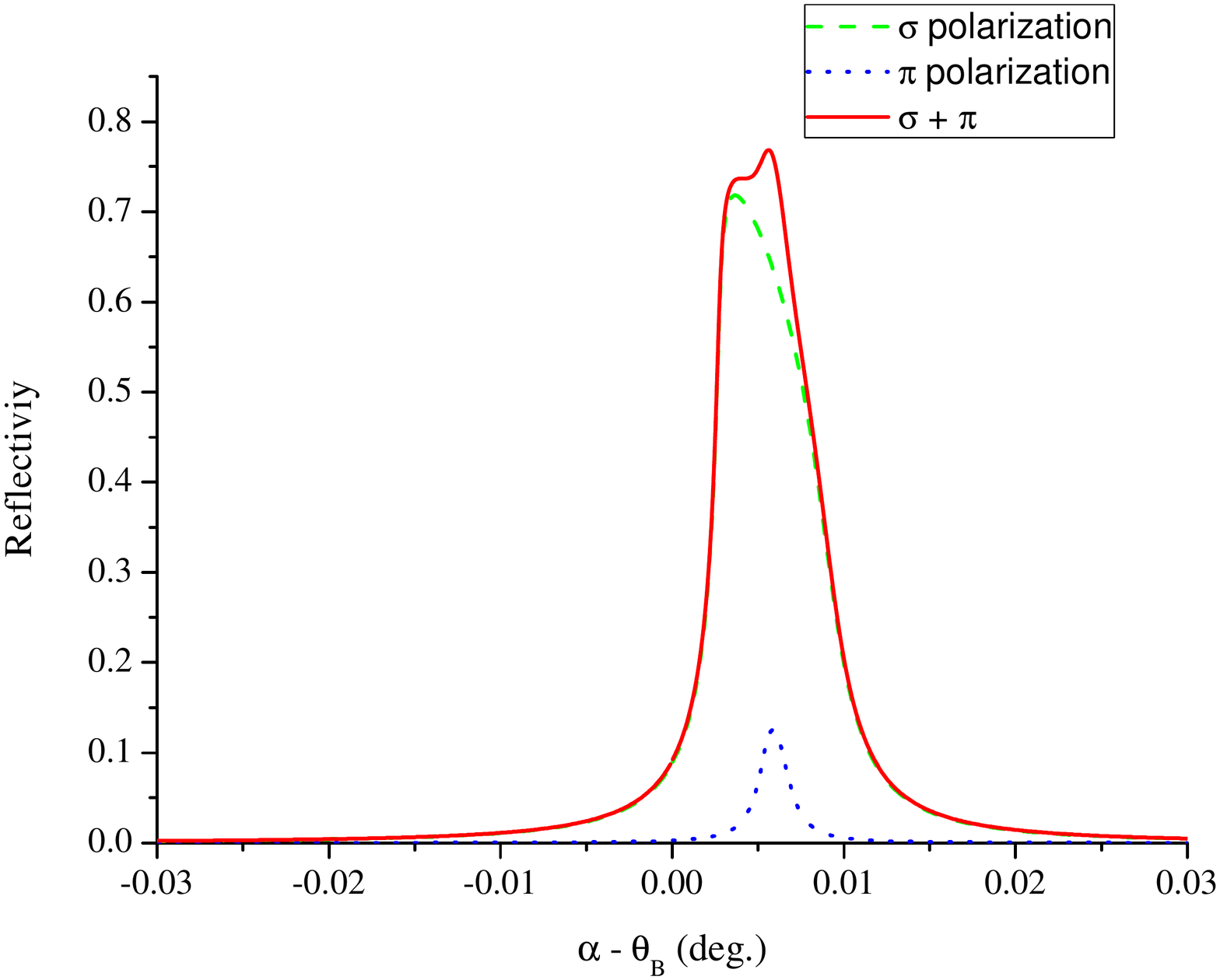}
\caption{(Color online) Si(111) reflectivity curve for
$\sigma$ (green dashed) and $\pi$ (blue dotted) polarizations as well as their sum (red full), evaluated with Xcrystal, a component of XOP.}
\label{fig:reflec}
\end{figure}

The  reflection on the crystals is described by dynamical diffraction theory. The reflectivity curve is created using the Xcrystal component of the XOP 2.3 program \cite{sad1998,sad2004} 
assuming an unpolarized x-ray source, and taking into account the reflections of both $\sigma$ and $\pi$ polarizations. Xcrystal implements the dynamical diffraction theory of Ref.~ \cite{zac1967}. This program has an input option for choosing between several input files with form factors obtained from different authors (Refs.  \cite{hgd1993,wak1995,cha1995,kis2000,cha2000,cha2011}). The  reflectivity curves are shown on 
Fig. \ref{fig:reflec} for a monochromatic line at an energy of \unit{3104}{eV}.
In the simulation program the reflectivity curve is interpolated using cubic splines and used as a probability distribution for the reflection of an x ray.
%

The reflectivity curve is evaluated at the Bragg angle corresponding to  the central wavelength $\lambda_1$. The curve depends on the index of refraction and absorption coefficients, which are energy-dependent. In the region of energy where we have performed  the measurements (\unit{3096}{eV} to \unit{3139}{eV}) the 
FWHM of the reflectivity curve changes 
by \unit{0.08}{\%/eV} and the peak reflectivity by \unit{0.1}{\%/eV}. The same diffraction profile is used for each  wavelength of an x-ray line distribution, since the variation of the diffraction profiles within the range of the peak is 
negligible (the typical widths of our lines are a few hundreds of meV FWHM).


The simulation program is also designed to take into account a small mosaicity of the crystal.  A Gaussian distribution for the orientation of the crystal surface ($\theta_{\mathrm{C}}$) is used.
When comparing experimental and simulated line profiles, we notice that this effect is very small and can be neglected. This  is consistent with the fact that the crystals had a special surface treatment as described in Sec.~\ref{subsec:crystals}. We also neglect the variation of the crystal lattice spacing as a function of position as it is measured to be small as described in Sec. \ref{subsec:crystals}.

Once the ray is reflected, the $y_a$ position along the $x_a y_a z_a$ axis is given by $y_a=-y''\sin(\theta_{\mathrm{T}}-\theta _{\mathrm{C}})$. The direction vector obtained from Eq.\eqref{eq:r2} is given in this axis by multiplying it by a rotation matrix along the $z$ axis with angle  $\theta_{\rm{T}}$.
The position vector at the second crystal in the non-dispersive or dispersive setup is obtained in the same way as for the first crystal with a translation given by Eq. \eqref{eq:trans} with  $L$ being the distance between the crystals and a projection over the surface of the second crystal. Similar to Eq. \eqref{eq:proj_sur}, the position at the second crystal crystallographic plane obtained after projection is given by
\begin{eqnarray}
y''_a&=&y'_a\frac{\cos(\theta)}{\cos(\pm \theta+\theta_{\mathrm{C}} \mp \beta)} ~ , \nonumber  \\
z''_a&=&z'_a\frac{\cos(\phi)}{\cos(\phi+\delta_2)} ~ ,
\label{eq:proj_sur_2}
\end{eqnarray}
where the plus and minus signs refer to the dispersive and non-dispersive modes, respectively. 
As in the case of the first crystal, if the ray position is inside the second crystal, then the glancing angle between the ray direction and the second crystal surface is calculated for the reflectivity. Furthermore, similar to the first crystal evaluation part, the glancing angle is obtained from Eq. \eqref{eq:alph} with $\bm{\hat{e}_a}$ defined after the first crystal reflection (Eq. \eqref{eq:r2}) and the normal vector of the second crystal $\bm{\hat{n}^\pm}_2$ is given by
\begin{eqnarray}
\hat{n}_{2x}^\pm    &=& \mp \cos(\delta_2) \sin(\theta_{\mathrm{T}} \pm \theta_{\mathrm{C}}+\beta) ~ , \nonumber  \\
\hat{n}_{2y }^\pm  &=& \pm \cos(\delta_2)  \cos(\theta_{\mathrm{T}} \pm \theta_{\mathrm{C}}+\beta)  ~ ,    \\
 \hat{n}_{2z}^\pm  &=&   \sin(\delta_2) \qquad\qquad\qquad\qquad~~~. \nonumber
\label{eq:vect_dire_2}
\end{eqnarray}
The direction vector of the reflected ray from the second crystal is given in
the $x_b y_b z_b$ coordinate system by multiplying it by a rotation matrix along the $z$ axis
with an angle $\theta_{\mathrm{T}}+\theta_{\mathrm{D}}^\pm$, where $\theta_{\mathrm{D}}^\pm$ is the
angle between the detector in the dispersive or non-dispersive modes and the axis of the source.
Finally, the position of the ray at the detector entrance plane in both modes is obtained with Eq. \eqref{eq:trans} with $L$ being the distance between the second crystal and the detector.
If the ray reaches the detector, then a count is added to the
simulated spectrum for the value of $\beta$.  

\section{Data Analysis}
\label{sec:data_an}


\begin{figure*}[tb]
\centering
\includegraphics[clip=true,width=\columnwidth,trim =0.5cm 3cm 0.5cm 0.9cm,]{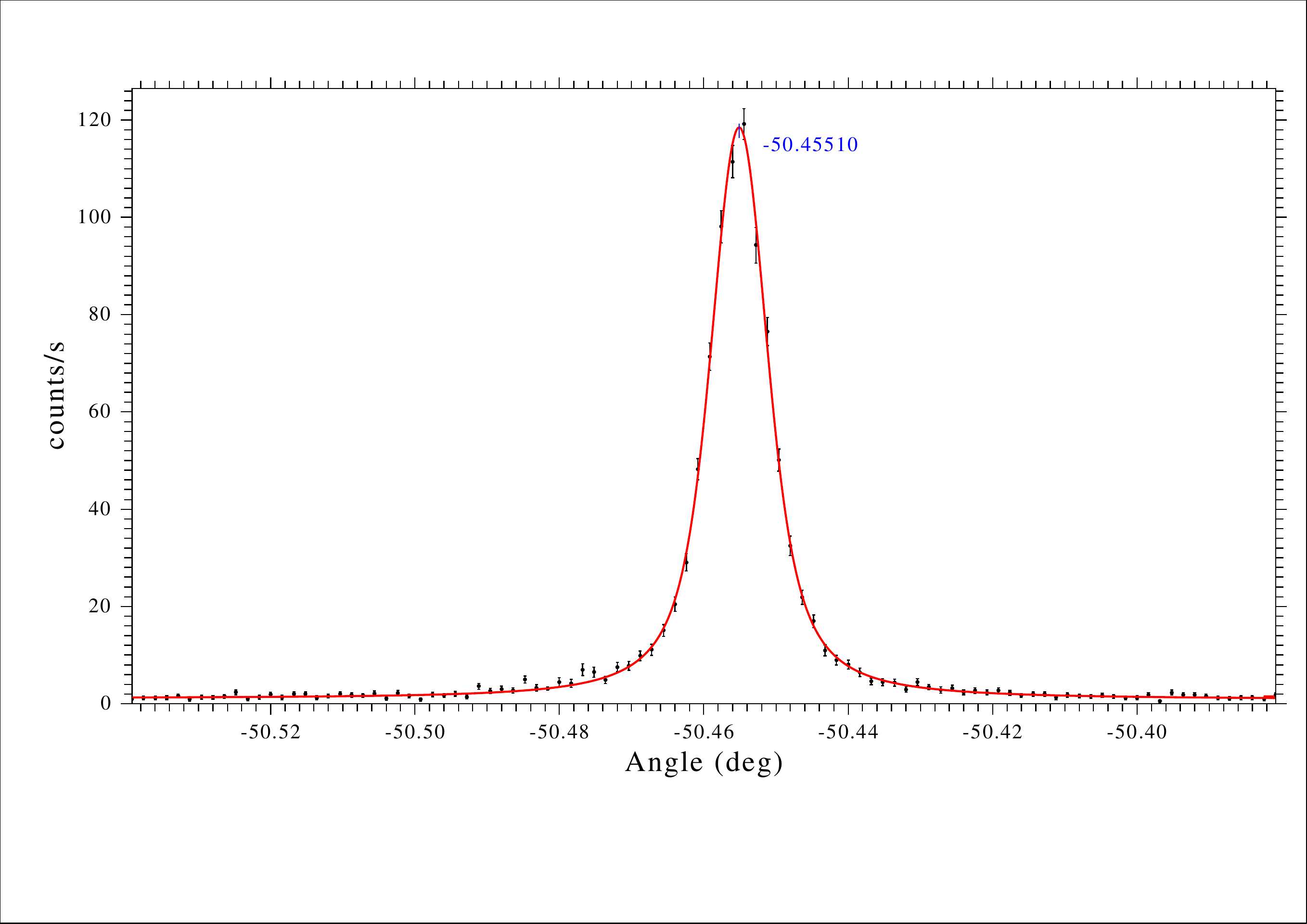}
\includegraphics[clip=true,width=\columnwidth,trim =0.5cm 3cm 0.5cm 0.9cm,]{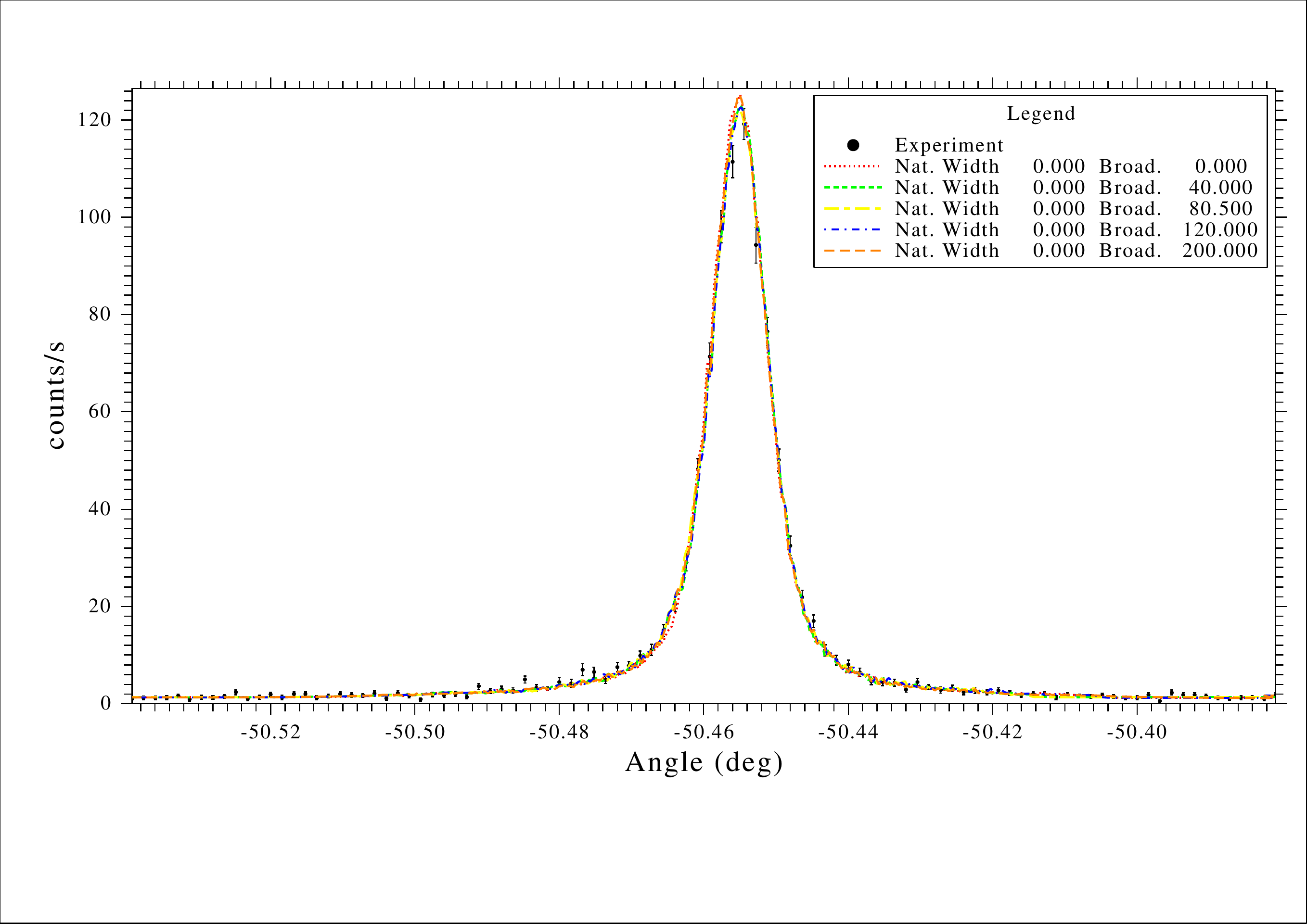}
\caption{(Color online) Voigt profile (left) and simulated profiles with different line width (right) fitted to an experimental non-dispersive mode spectrum. The Voigt profile fit yield a $\chi^2=1.12$ while the simulated ones  gives $\chi^2$ ranging from 1.2 to 1.4 (the variation is mostly due to the statistic of the simulated profiles). The difference in angle is equal to \unit{1.3\times 10^{-4}}{degrees}.}
\label{fig:parallel-fit-comp}
\end{figure*}

\begin{figure*}[tb]
\centering
\includegraphics[clip=true,width=\columnwidth,trim =0.5cm 3cm 0.5cm 0.9cm,]{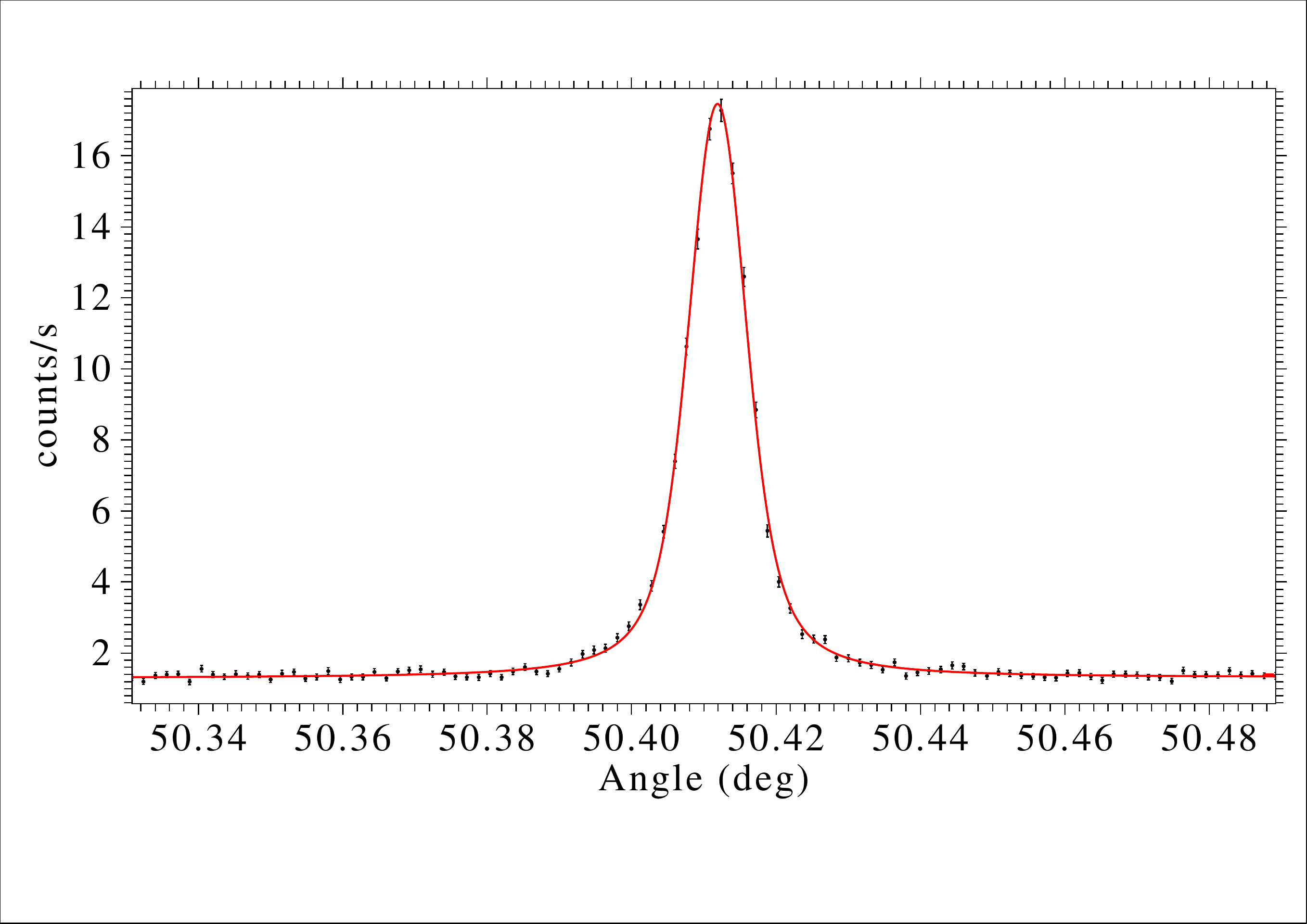}
\includegraphics[clip=true,width=\columnwidth,trim =0.5cm 3cm 0.5cm 0.9cm,]{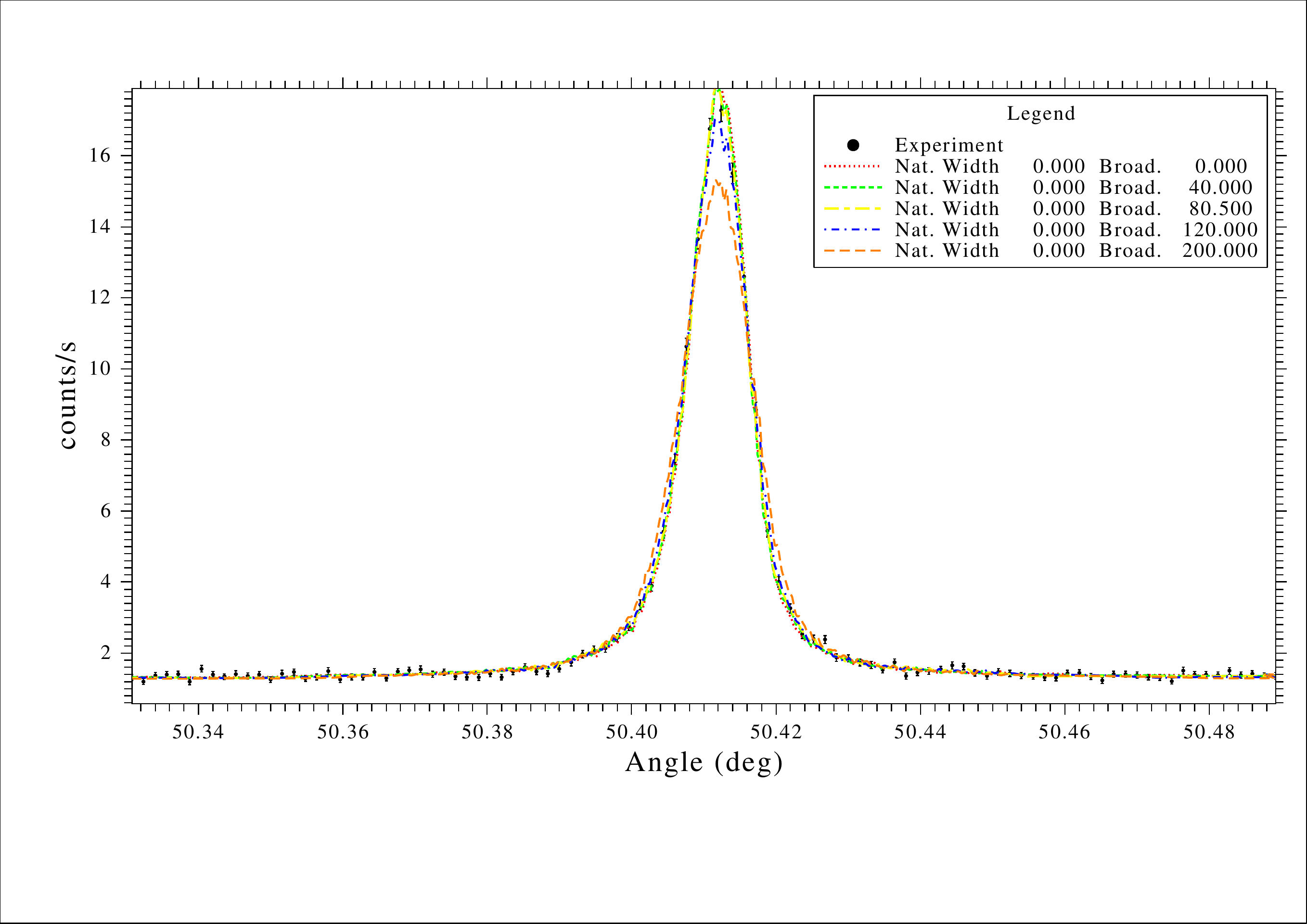}
\caption{(Color online) Voigt profile (left) and simulated profiles with different line widths (right) fitted to an experimental dispersive mode spectrum. The Voigt profile fit yield a $\chi^2= 1.42$ and the optimum simulated profile one (for a Gaussian broadening of \unit{80}{meV}) gives a $\chi^2=1.26$. The difference in angle  between the peak position between the two fits is equal to \unit{1.86(7)\times 10^{-3}}{degrees}.}
\label{fig:dispersive-fit-comp}
\end{figure*}

In previous work, the method to deduce the energy from experiment was to fit the experimental spectrum with a Voigt profile to obtain the angle position of the peak and apply the Bragg law with index of refraction and vertical divergence corrections.
The problem is that the crystal reflection curve is asymmetric (Fig. \ref{fig:reflec}). In non-dispersive mode, the asymmetry disappears because the rocking curve is the convolution of the profile of Fig.  \ref{fig:reflec} convolved with its mirror image. Figure \ref{fig:parallel-fit-comp} shows an experimental non-dispersive spectrum  fitted with a Voigt profile and with a sequence of simulations, corresponding to different Gaussian broadenings of the x-ray line. It shows clearly that within the statistical uncertainty of the simulated profile the fit quality is the same as with a Voigt profile. In addition, Fig.  \ref{fig:parallel-fit-comp} clearly demonstrates the high level of agreement between  the simulation and the experimental profile. We would like to emphasize that, except for the energy used in the simulation, there is no adjustable parameter here. Moreover, since it is a non-dispersive profile, the width of the line does not change the shape as expected. 
In the dispersive side, the reflection curve is convolved with itself, which enhances the asymmetry. An example of a fit by a Voigt and fits by a sequence of simulated profiles is shown in Fig. \ref{fig:dispersive-fit-comp}. Again, for the profile with the width that provides the smaller $\chi^2$, the fit quality is excellent, and the reduced $\chi^2$ very close to 1, showing the quality of the simulation. The asymmetry of the line translates into a 
difference of   \unit{1.86\times 10^{-3}}{degrees.} between the peak positions obtained from the simulation and the Voigt fit, while it is only \unit{1.3\times 10^{-4}}{degrees} in the non-dispersive side. In the dispersive side, it corresponds to 19 times the angular encoder error. 
Moreover, because of the complicated line shape, the value of the angle corresponding to the peak position of the simulated profile itself is not a well defined quantity. The only well defined quantity is the energy of the line that has been used in the simulation.
To avoid this problem, we used two methods. In the first one we used an analytic approximation of the profile, which allowed to have a direct relation between the energy and the peak position.  \cite{ama2011} In a second one, we fitted both the simulated and the experimental profile with a Voigt profile. The difference in angle $\delta \theta$ and in temperature $\delta T$ are used to correct the energy used in the simulation.

We write the line energy as
\begin{eqnarray}
E(T,\theta)&=&\frac{C n}{2d\left(1+\alpha\left(T-T_0\right)\right)} \nonumber \\ 
&&\times \frac{1}{\sin\left(\theta+\chi \tan \theta\right)\left(1-\frac{\delta}{\left(\sin \theta\right)^2}\right)}
\label{eq:E-exp}
\end{eqnarray}
where $n$ is the order of diffraction,  $\delta$ the index of refraction, $C=h c$ is the wavelength to energy conversion factor equal to \unit{12398.41875(31)}{eV}\AA \cite{mtn2008}.
The coefficient $\chi$ is the vertical divergence correction
\begin{equation}
\chi=\frac{a^2+b^2}{24L^2},
\end{equation}
where $L$ is the distance between the slits which defines the height of the spectrometer (in our case the polarization electrode and the detector window), and $L$ the distance between these slits.
The final energy $E_f$ is written in term of the simulation energy $E_s$ as
\begin{equation}
E_f=E_s + \frac{\partial E(T,\theta)}{\partial T}\delta T +  \frac{\partial E(T,\theta)}{\partial \theta}\delta \theta
\end{equation}

The fit program uses the least-square method , with the Levenberg-Marquardt algorithm, in the implementation of Ref.  \cite{pftv1986}. 

The Voigt profile is a convolution product of a Lorentzian (representing the emission profile of the line) and of a Gaussian (representing an instrumental broadening), see, e.g., Ref.  \cite{arm1967}. 
It is written as
\begin{equation}
I(\theta,\theta_0,\ell,g)=I_0 \frac{K(x,y)}{K(0,y)}
\label{eq:line-profile}
\end{equation}
with the reduced Voigt function
\begin{eqnarray}
K(x,y)&=&\frac{y}{\pi}\int_{-\infty}^{\infty}dt\frac{e^{-t^2}}{(t-x)^2+y^2} \\
x&=& \frac{2(\theta-\theta_0)\sqrt{\ln 2}}{g}\\
y&=&\frac{\ell}{g}\sqrt{\ln 2},
\label{eq:line-voigt-func}
\end{eqnarray}
where $\theta_0$ is the peak position, $I_0$ the intensity at $\theta_0$, $\ell$ the Lorentzian FWHM and $g$ the Gaussian FWHM.
The FWMH of the Voigt profile can be given to a very good approximation as:
\begin{equation}
w=\frac{\ell+\sqrt{\ell^2+4g^2}}{2}.
\label{eq:total-voigt-width}
\end{equation}
An exact expression was derived in Ref.  \cite{jaq2007}, Eq. (21). It provides values in excellent agreement with the previous equation, but is much less convenient to use.
The Voigt profile and the needed derivatives are evaluated following the method described in Ref.  \cite{arm1967,ind1983}.

%

%
%
%
%

%
\section{Study of Uncertainties}
\label{sec:uncertain}

The systematics errors in the measurement performed using the spectrometer described here can be divided into three categories. The first one includes  the uncertainties due to the alignment and to the precision of the construction of the DCS.  The second one is related to uncertainties in the knowledge of 
 diffraction profiles and on the polarization of the x~rays. The third category is due to the uncertainty of the knowledge of fundamental constants or crystal properties like the lattice spacing. 
 Uncertainties from the  first  two categories can be estimated with the help of the simulation program described in Sec. \ref{sec:theoryDCS}. The energy deduced from simulated spectra, with various parameters varied, is evaluated following the method of Sec.~\ref{sec:data_an} and compared with the simulation input energy. We give in Table \ref{tab:lis_uncer} the list of contributions to the final error budget for the absolute measurement of the He-like Ar $1s2s\,^3S_1 \to 1s^2 \,^1S_0$ M1 transition at \unit{3104.148}{eV} \cite{asyp2005}. Most contributions to the uncertainty change very slowly with energy. The different contributions are explained below.

\subsection{Geometrical uncertainties}
\label{subsec:geo_unce}

\begin{figure}[tb]
\centering
\includegraphics[clip=true,width=\columnwidth]{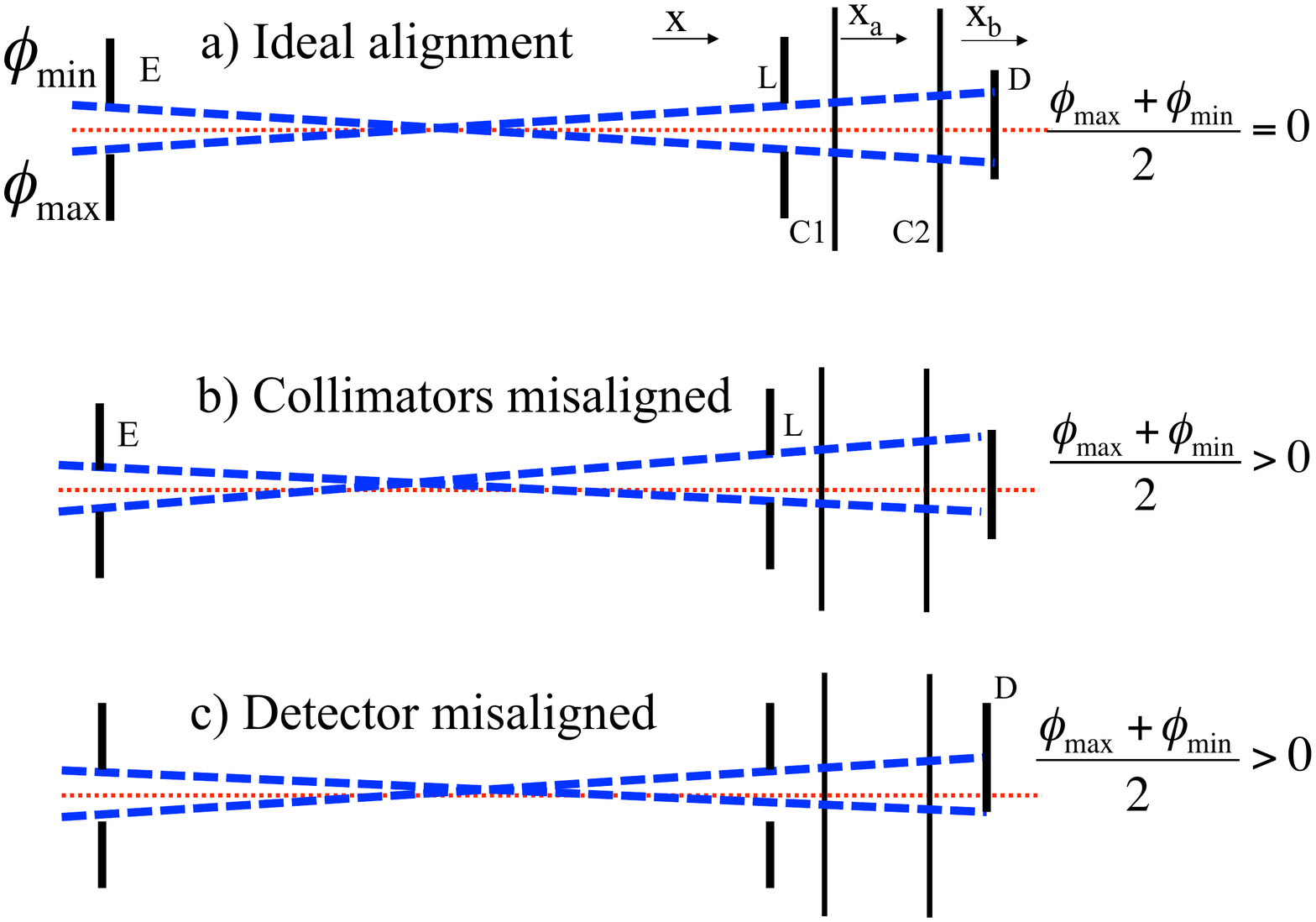}
\caption{(Color online)Scheme of the collimator system. The points E and L  refers to the entrance of the x~rays and to the lead collimator respectively, as represented in Fig.~\ref{fig:simpa-DCS-geom}.  C1, C2 and D represents the first and second crystal and the detector, respectively.
 Figure a) represents an ideal alignment;  Fig. b) a vertical misalignment of L compared to E;   Fig. c)  a vertical misalignment of the detector.  The  dashed lines represent rays with either maximum or minimum vertical divergence $\phi$. The  dotted line is the symmetry axis.}%
\label{fig:collim_sque}%
\end{figure}

\begin{figure}[tb]
\centering
\includegraphics[clip=true,width=\columnwidth,trim =0.5cm 1cm 0.5cm 0.9cm]{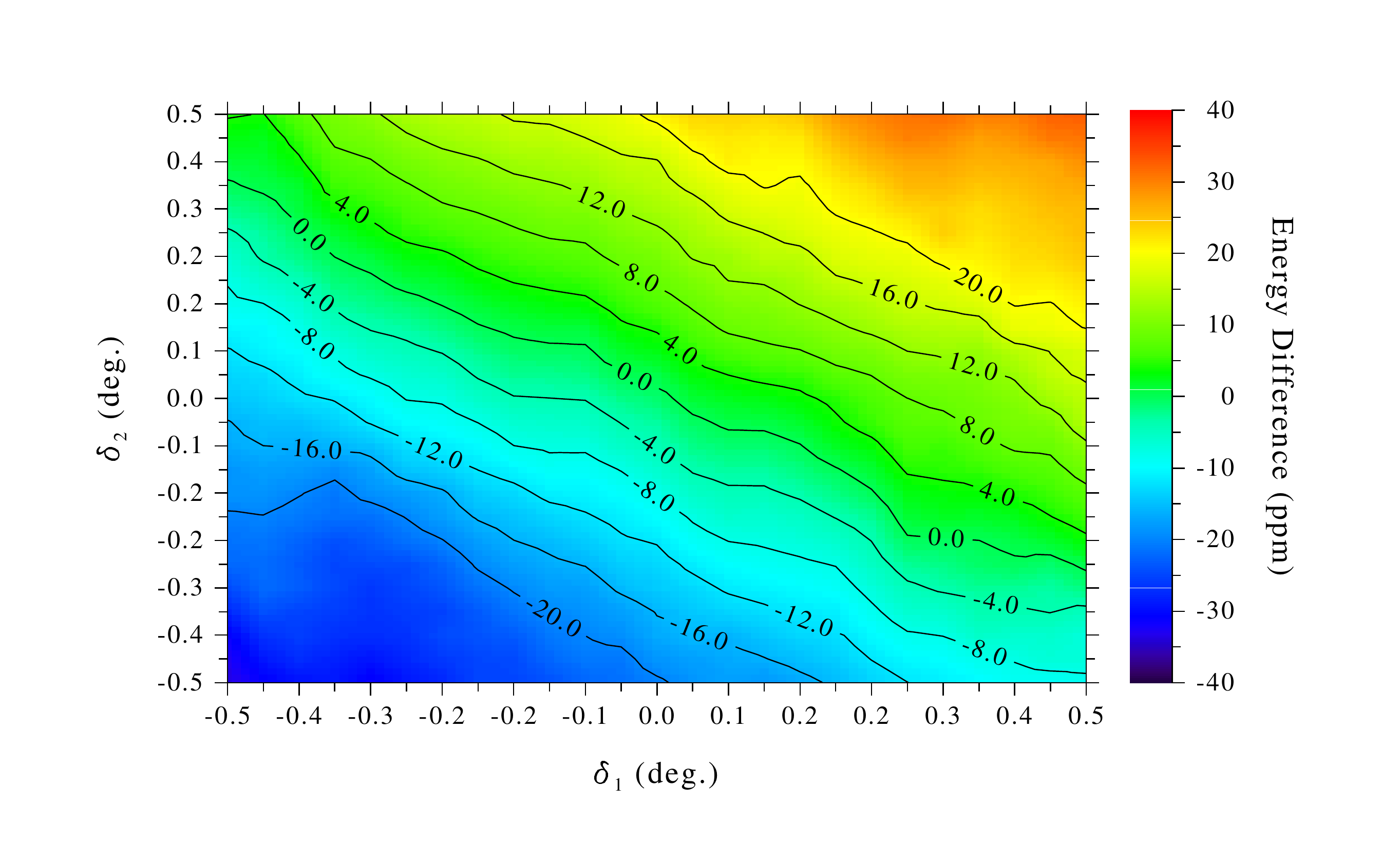}
\caption{Plot of the energy difference (ppm) between upper and lower masks used on the second crystal for several values of crystal tilts of $\delta_{1}$ and $\delta_{2}$. (Color online.)}%
\label{fig:dif_energ_masks}%
\end{figure}

\begin{figure}
[tb]
\centering
\includegraphics[clip=true,width=\columnwidth]{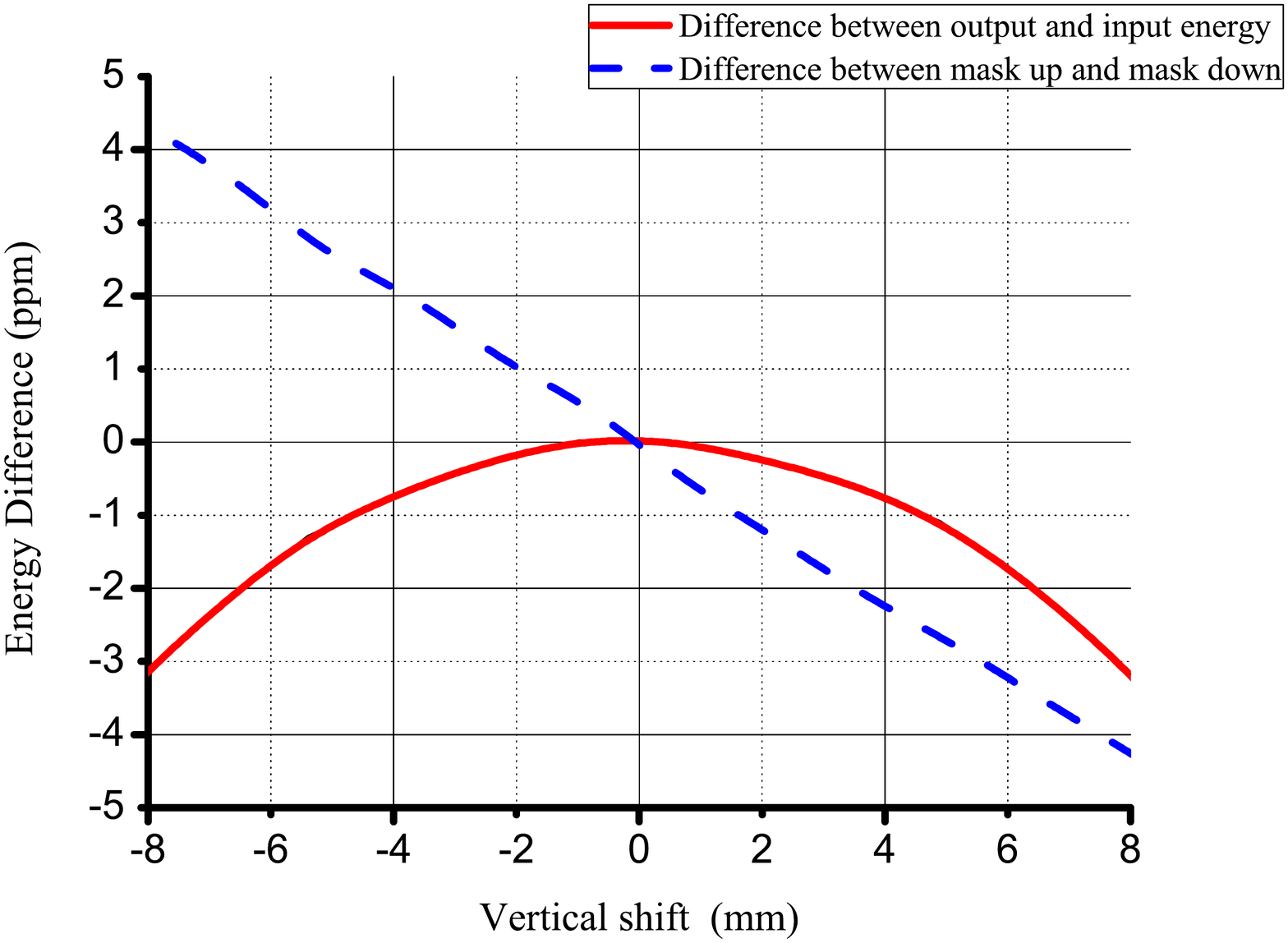}
\caption{(Color online) Energy difference (ppm) between configurations with an upper or lower mask for several values of a vertical shift in the position of the spectrometer's collimator. The solid line is the difference between the input and output energies of the simulation. The dashed line is the difference between the simulation energy outputs for the upper and lower masks.}
\label{fig:dif_energ_masks_verMis}%
\end{figure}

The first two geometrical uncertainties are related to the alignment. The uncertainty in the verticality of the crystal diffracting planes is due to the alignment procedure described in Sec. \ref{subsec:align} and to the error in the cut angle of the crystal (Sec. \ref{subsec:crystals}). 
To this uncertainty, one has to add the one due to a possible misalignement of the DCS input collimators (Fig. \ref{fig:collim_sque}).
The total effect of these misalignments  can been checked by recording x-ray spectra with absorbing masks that cover alternatively the upper and lower halves of the crystals. The comparison of the energies obtained in the two measurements gives an indication of the total uncertainty on the alignment within the statistical uncertainty of the measurement. Figure \ref{fig:dif_energ_masks} shows the simulated energy difference obtained
with upper and lower mask positions for several values of crystal tilts, $\delta_{1,2}$.
Similarly, Fig.~\ref{fig:dif_energ_masks_verMis}  shows the energy
difference between the upper and lower mask cases for several values of vertical shifts of the lead collimator (see Fig.~\ref{fig:collim_sque} b) that connects
the source to the spectrometer.

As explained in Sec.  \ref{subsec:align}, the alignment procedure provides \unit{\delta_{i}\le 0.01}{ degrees}. The uncertainty related to crystal tilts was obtained from the simulation program, comparing energies from simulations using \unit{\delta_i=0, \pm 0.01}{degrees}. This uncertainty  is in good agreement with the expressions of Bearden and Thomsen \cite{bat1971}.

The uncertainty related to the vertical misalignment of collimators was obtained in a similar way by running simulations with a collimator entrance shifted by \unit{\pm 0.45}{mm} (see Fig.~\ref{fig:collim_sque} a and b), i.e., with a vertical shift of the collimator so that \unit{(\phi_{\mathrm{max}}+\phi_{\mathrm{min}})/2=\pm0.01}{degrees} (the total spectrometer length is \unit{2.6}{m}). The relevant dimensions are given in Fig.~\ref{fig:simpa-DCS-geom}). The equivalent situation for   a vertical shift in the detector position is represented in Fig.~\ref{fig:collim_sque} c). From a geometrical point of view, it is irrelevant which elements are restricting $\phi_{\mathrm{max}}$ and $\phi_{\mathrm{min}}$. We thus performed a single simulation, shifting the input collimator by \unit{\pm 1}{mm}, leading to a large overestimate of the total uncertainty.

\begingroup
\begin{table*}[tb]
\begin{center}
\begin{tabular}{ld}
							
	Contribution	&	\multicolumn{1}{c}{Value (\unit{eV})}	\\
\hline
			\multicolumn{2}{c}{Geometrical uncertainties}				\\
Crystal tilts ($\pm$ 0.01° for each crystal)	&	0.0002	\\
Vertical misalignment of collimators (1 mm)	&	0.0002	\\
X-ray source size (6 to \unit{12}{mm})	&	0.0013	\\
\hline							
		\multicolumn{2}{c}{Diffraction profile uncertainties}			\\
Form factors 	&	0.0020	\\
X-ray polarization	&	0.0014	\\
\hline							
		\multicolumn{2}{c}{Instrumental limitations and uncertainties on physical constants}			\\
Fit and  extrapolation to standard temperature	&	0.0044	\\
Angle encoder error	(\unit{0.2}{arcseconds})&	0.0036	\\
Lattice spacing error	&	0.0001	\\
Index of refraction	&	0.0016	\\
Coefficient of thermal expansion	&	0.0002	\\
Temperature (\unit{0.5}{°C})	&	0.0040	\\
Energy-wavelength correction	&	0.0001	\\
\hline
Total	&	0.0077	\\
\end{tabular}
\end{center}
\caption{\label{tab:lis_uncer}
List of uncertainties and error contributions. The simulations were performed for an x-ray energy  of \unit{3104.148}{eV}, which corresponds to the $1s2s \,^3S_1 \to 1s^2 \,^1S_0$ transition in He-like argon. The uncertainty due to form factors was obtained by
comparing simulations with different form factor values from Refs.  \cite{hgd1993,wak1995,cha1995,kis2000,cha2000,cha2011}. The x-ray polarization uncertainty is obtained by comparing a simulation done with a crystal reflection profile for a fully $\sigma$-polarized  and an unpolarized beam.}
\end{table*}
\endgroup

%
%

%
 %
%
%


Previous measurements with a DCS used high-power x-ray tubes to provide x-ray lines from solid targets or sometimes gas targets \cite{mlik1992}. Here we use a plasma, the geometry and position of which are fixed by the field configuration, the microwave frequency and power, and possibly other factors like ionic and electronic temperatures and polarization electrode bias. The plasma, as fixed by the magnetic field structure, is \unit{\approx30}{mm} in diameter. Yet, x-ray imaging was performed before on ECRIS \cite{bvst2004}, which shows that the HCI position with respect to the source axis may change depending on the operation conditions. To estimate possible uncertainties due to this effect, we performed two simulations for an x-ray plasma diameter of \unit{12}{mm} (diameter of the collimator) and another for a \unit{6}{mm} plasma diameter.  We find a difference of \unit{1.3}{meV}, which we use as a largely over-estimated uncertainty in Table \ref{tab:lis_uncer}.

Besides vertical and horizontal angle shifts, the case of the alignment uncertainty due to a possible vertical or horizontal translations of the crystals was also considered. No observable difference was noticed in the simulated results.

\begin{figure}
[tb]
\centering
\includegraphics[clip=true,width=\columnwidth,trim =0.2cm 0.2cm 0.2cm 0.2cm]{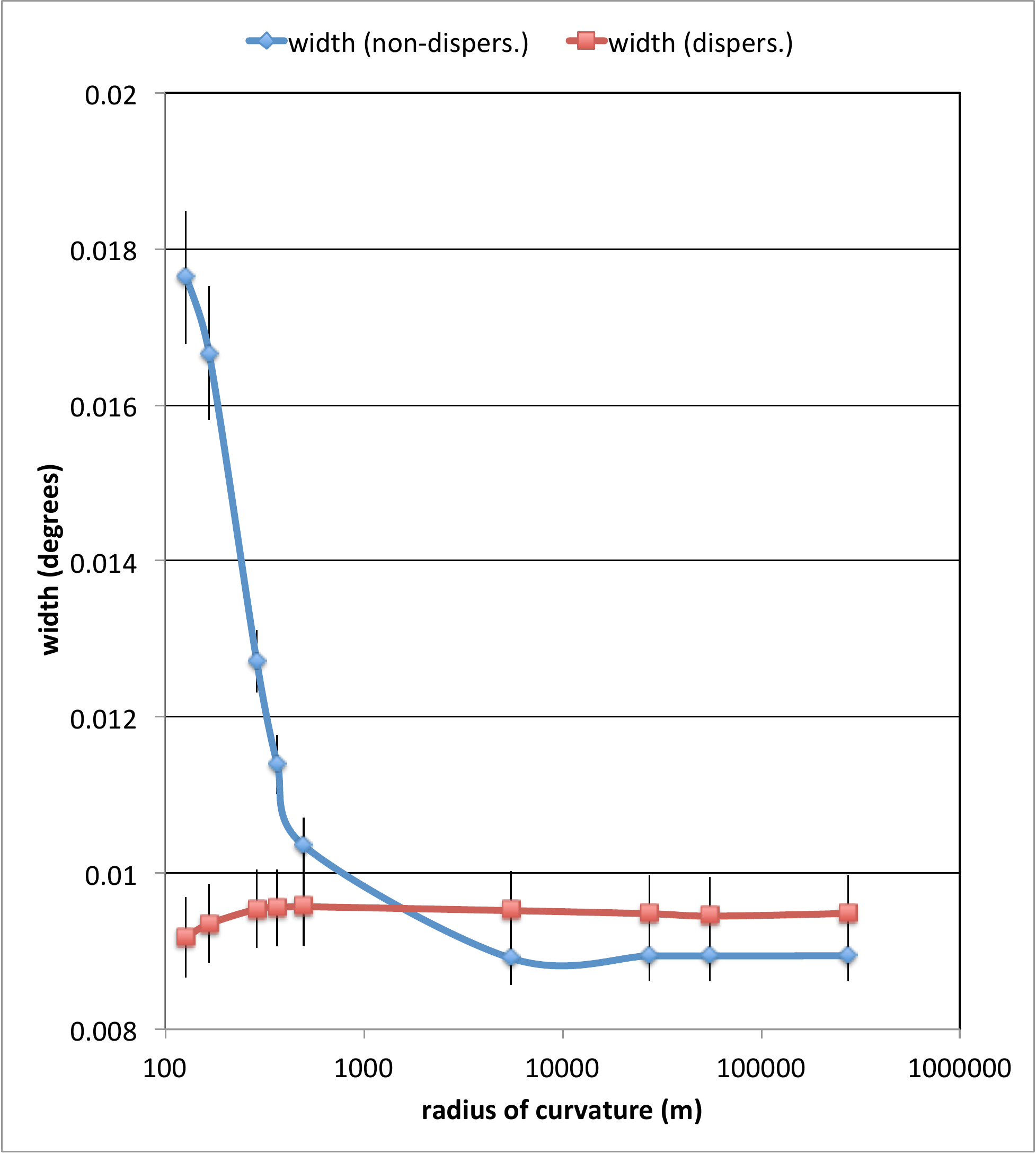}
\caption{(Color online)Widths of the dispersive and non-dispersive profiles as a function of the crystals radius of curvature. These widths are obtained by fitting simulated spectra with a Voigt profile and combining the Lorentzian and Gaussian widths using Eq. \eqref{eq:total-voigt-width}. Error bars are due to statistics. The dispersive and non-dispersive widths are identical for large radii of curvature as expected, within simulation statistical uncertainty.\label{fig:widths-curv}}%
\end{figure}

\begin{figure}
[tb]
\centering
\includegraphics[clip=true,width=\columnwidth,trim =0.2cm 0.2cm 0.2cm 0.2cm]{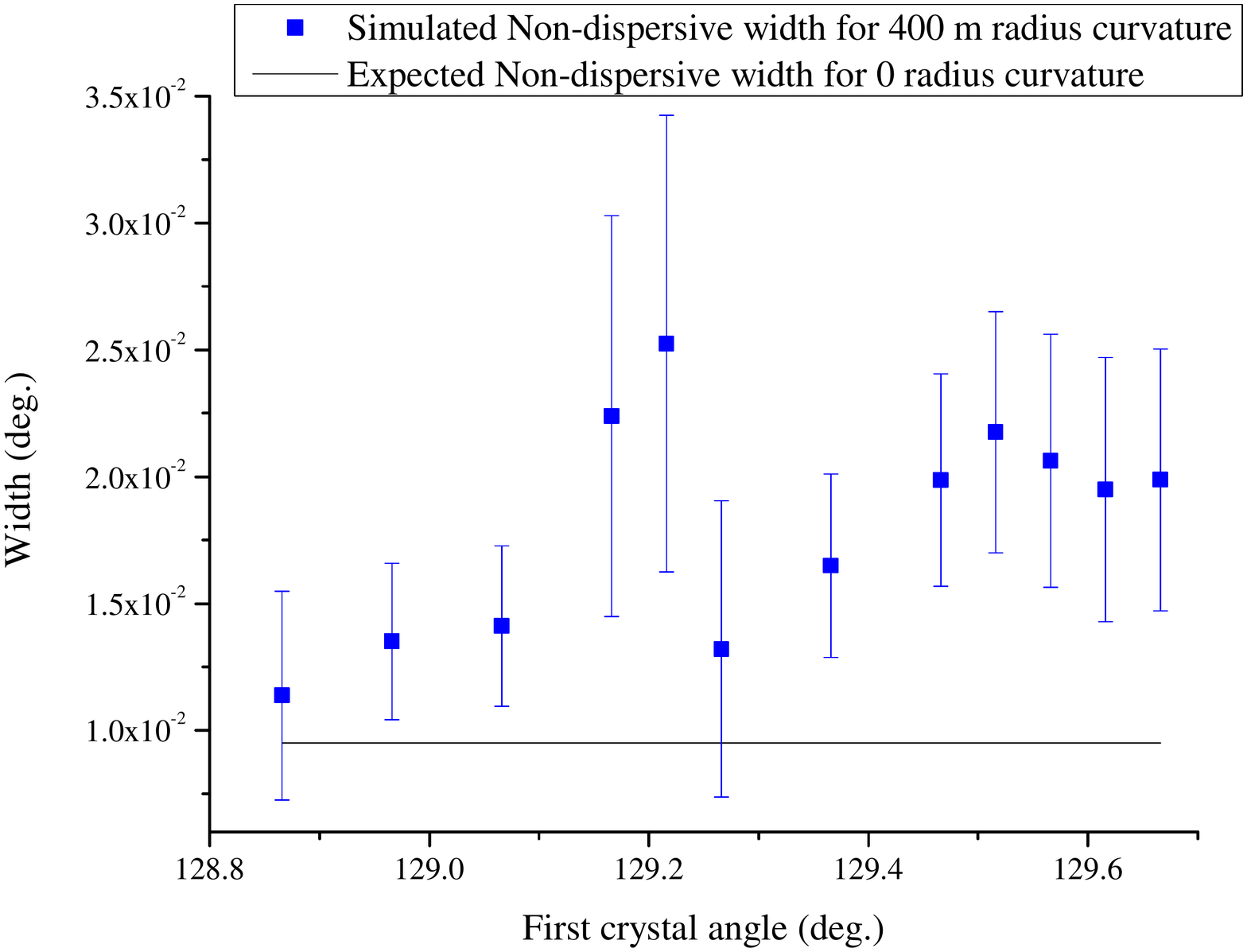}
\caption{(Color online) Simulations performed for several values of the first crystal angle and a curvature radius of \unit{400}{m} in both crystals for the He-like Ar M1 line.  The simulated non-dispersive width is plotted for several values of the first crystal angle.\label{fig:first_widths}}%
\end{figure}

\begin{figure}
[tb]
\centering
\includegraphics[clip=true,width=\columnwidth,trim =0.2cm 0.2cm 0.2cm 0.2cm]{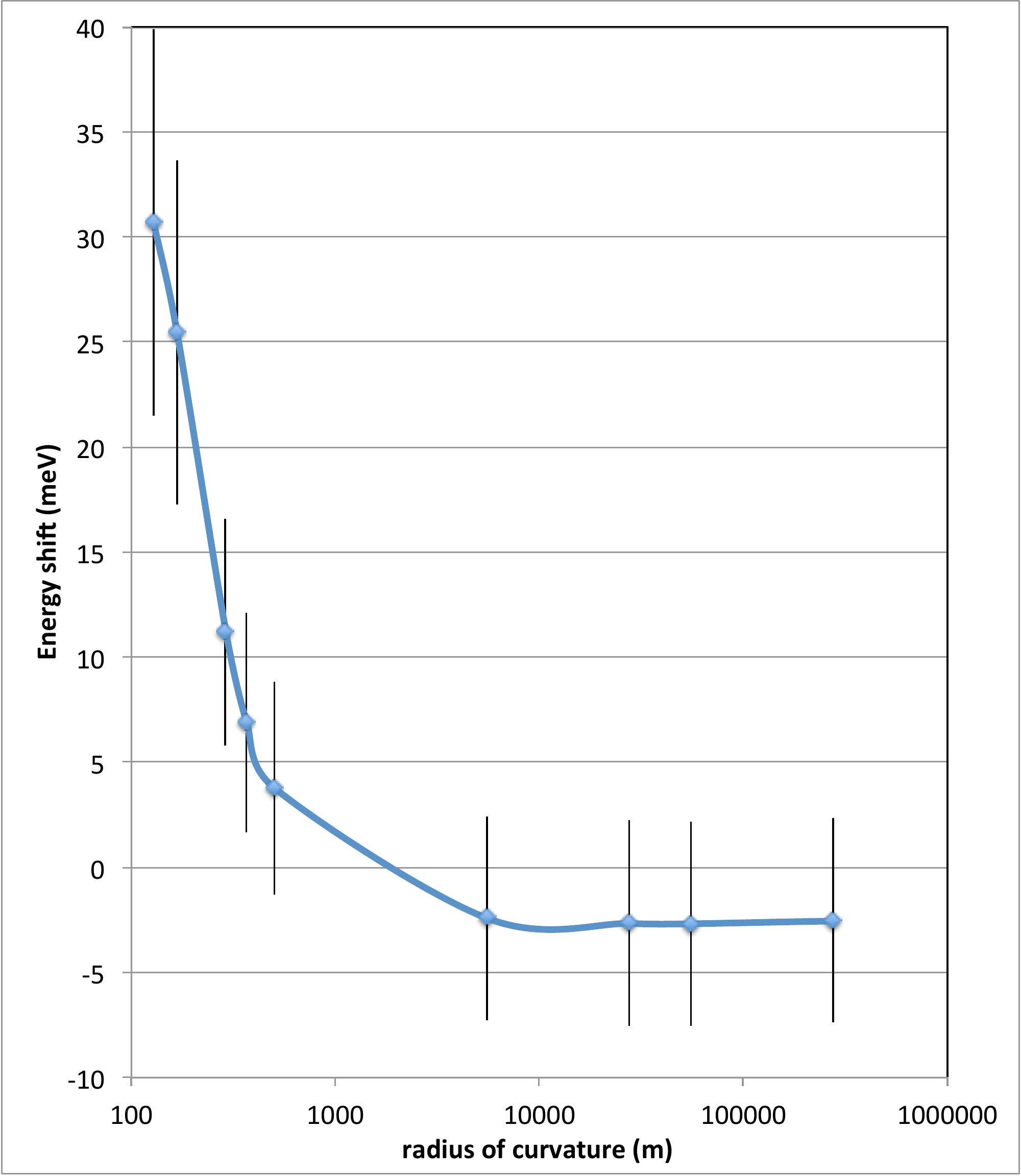}
\caption{(Color online)Energy shift due to a curvature of the two crystals. These shifts are obtained by fitting simulated spectra with a Voigt profile. Error bars are statistical error bars.\label{fig:energy-curv}}%
\end{figure}

Another possible uncertainty source could be due to a small crystal curvature. Simulations performed for this effect show that  the non-dispersive profile is the most sensitive to curvature. The variation of the width of the dispersive and non-dispersive profiles as a funciton of the radius of curvature are shown in Fig. \ref{fig:widths-curv}. Changes in the dispersive side width are small, at the limit of the statistical significance. Changes in the non-dispersive side are large for radii of curvature smaller than \unit{\approx 1000}{m}. 
The crystal curvature also induces a dependence of the non-dispersive spectra width on the first crystal angle as can be seen in Fig.~\ref{fig:first_widths}.  Finally the dependence of the line energy on the radius of curvature is plotted in Fig. \ref{fig:energy-curv}. The figure shows that for radii of curvature larger than \unit{\approx 5000}{m} the shift is much smaller than the statistical error on the fit.
This effect is experimentally minimized by using thick crystals (\unit{6}{mm}) and nylon screws  just brought to contact, to hold the crystal against the reference surface of the support as described in section \ref{subsec:dcs}. We are able to see experimental evidence of crystal bending when pressing them hard against their supports with strongly tightened brass screws. We were then able to observe experimentally a broadened  line profile in the non-dispersive mode, corresponding to a bending radius of \unit{\approx 500}{m} and a dependence of the width on the first crystal angle as reported in Fig. ~\ref{fig:first_widths}. This effect disappeared completely with the normal mode of holding the crystals, and the parallel profiles show no signs of broadening as seen in Fig. \ref{fig:parallel-fit-comp}.

%
%

%

%
%
%

\subsection{Diffraction profile uncertainties}
\label{subsec:dif_unce}

The energy values obtained with a DCS in reflection at low energy depend critically on the quality of the crystal reflectivity curve. The peak position is dependent on the index of refraction, for which  few experimental determinations exists, all obtained  at high energy.
As described in Sec. \ref{sec:theoryDCS} we use  two programs , XOP \cite{sad2004,sad2004a,sad1998} and X0h, \cite{las1991,stepanov} to calculate  reflectivity curves in the simulations. Moreover we use the capacity of 
XOP  to  choose different form factor values \cite{hgd1993,wak1995,cha1995,kis2000,cha2000,cha2011}. By comparing  simulations performed with the diffraction curves from the two different programs and with the different form factors, we obtain an uncertainty of \unit{2}{meV} for the diffraction profile.

The index of refraction provided by XOP is $5.1005\times 10^{-5}$ for a line energy of \unit{3104.148}{eV}. Henke et al.  \cite{hgd1993} provide the semi-empirical value of $5.0790\times 10^{-5}$, and Brennan and Cowan \cite{bac1992} value is $5.0825\times 10^{-5}$. The maximum variation of
the final energy using the different values of the index of refraction is \unit{1.6}{meV}, which we use as an uncertainty.

The uncertainty due to unknown \emph{polarization} of the x rays was also estimated with the use of simulations. We performed two simulations; one with a diffraction profile containing only the $\sigma$ polarization and another with $\sigma+\pi$ polarization (unpolarized). From the difference a maximum uncertainty of \unit{1.4}{meV} can be estimated due to the presence of any polarized light. The integrated reflectivity using only $\pi$ polarization is \unit{6}{\%} of the the one obtained with $\sigma$ polarization. This would lead to roughly 230 times fewer counts.  The width of the profile obtained using only $\pi$ polarized x rays  is roughly \unit{30}{ \%} smaller than the width of a profile obtained with the $\sigma$ polarization. The agreement between experimental profile widths and simulation widths performed for unpolarized  x rays is excellent. This confirms within the statistical uncertainty in the experimental spectra that the x-ray beam from the ECRIS is not polarized and justify the uncertainty we quote in Table \ref{tab:lis_uncer}.

\begin{figure}
[tb]
\centering
\includegraphics[clip=true,width=\columnwidth]{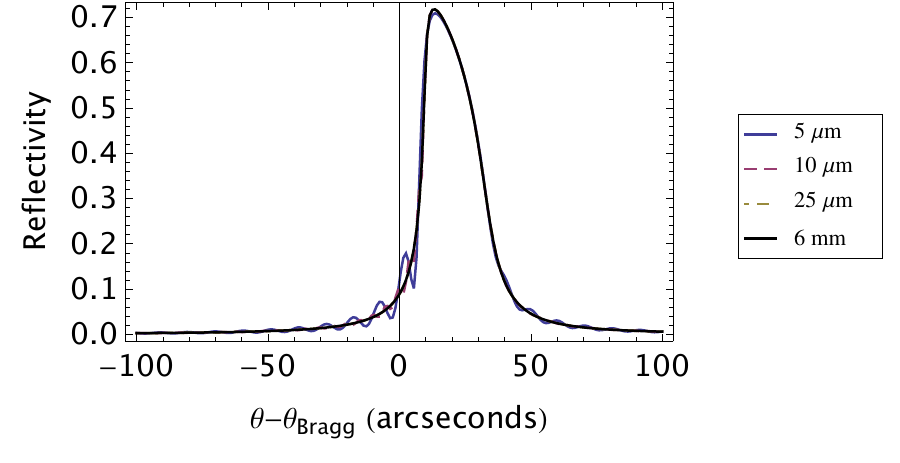}
\caption{(Color online)Reflectivity curves for different crystal thicknesses evaluated with XOP. Only the \unit{5}{\mu m} and  the \unit{10}{\mu m} profiles shows \emph{pendellösung} oscillations.\label{fig:pendell}}%
\end{figure}

We also considered other effects, like distortion  of the diffraction profile due to \emph{pendellösung}. The changes in the crystals diffraction profile at the He-like Ar M1 transition energy are presented  on Fig.  \ref{fig:pendell}. These effects are completely washed out for crystal thicknesses above \unit{20}{\mu m}, while our crystals have a thickness of \unit{6}{mm}. The same result was obtained with both XOP and X0h.

The penetration depth of \unit{3.1}{keV} x rays in Si is very small. We get \unit{1.44}{\mu m} for $\sigma$-polarization with XOP. This corresponds to an extinction length of \unit{2.26}{\mu m}. For the $\pi$-polarization they are \unit{7.6}{\mu m} and \unit{12}{\mu m} respectively, but we have seen that this polarization contributes only a small fraction of the profile. We take into account the fact that each ray is reflected at a different depth in the crystal in the simulation. An exponential distribution with mean value of the \emph{penetration depth}  is used to  obtain  the depth where the ray is reflected.  Simulations show that it would require  penetration depths  of a few mm to change significantly the measured energy. This effect can thus be  completely neglected.

The effect of a small \emph{mosaicity} of the crystals was also considered as a source for possible broadening of the diffraction profiles. Simulations show that non-negligible uncertainties due to this effect can only happen for values of mosaicity that produce a much larger width of simulated non-dispersive profiles than the ones observed in the experiment.  

The method of alignment of the crystals for polishing gives rise to an \emph{asymmetric cut} uncertainty of \unit{\approx 10}{arcseconds} (Sec. \ref{subsubsec:crystal-prep}). We used XOP to estimate a possible broadening of the diffraction profile due to this and no difference was observed. We simulated the effect  of an asymmetric cut  of that size on the energy and found none.This uncertainty was checked experimentally by turning the crystal upside down between two measurements and by comparing the non-dispersive profiles. With the experimental statistics reachable in the current setup, we did not observe any difference in the diffraction profiles with flipped crystals. This gives us confidence on the present uncertainty due to the asymmetric cut of the crystals.


\subsection{Other sources of uncertainty}
\label{subsec:dif_ana_unc}

 As can be seen from Table \ref{tab:lis_uncer}, the largest source of uncertainty  comes from the statistical uncertainty of the fit,  and from the extrapolation of data taken at different temperature to standard temperature (\unit{22.5}{°C}). To this must be added uncertainties on fundamental constants and 
 crystal physical properties. 
 
 The main source of uncertainty lies in the difficulty of stabilizing the crystal temperatures under vacuum, with the stepping motors heating the crystal supports and the ECRIS klystron heating the room. The temperature controller is perfectly able to maintain a very stable temperature at atmospheric pressure, but not under vacuum. The use of special graphite contact sheets to improve the contact between the thermistors and the crystal could not completely fix the problem. Most of the time, it was not possible to set the temperature to below \unit{22.7}{°C}.  In order to alleviate this difficulty, we perform sequences of measurements at various temperatures and extrapolate to \unit{22.5}{°C}. This problem leads then to two different uncertainties: one is due to the precision of the temperature measurement, which we assume to be much worse than the calibration of the thermistors. The second is due to the extrapolation procedure, which combines the statistical uncertainty of the peak position determination and the one due to the fit of a linear function to the temperature dependance of the peak positions and extrapolation to standard temperature.
This problem will be fixed in the next version of the crystal supports, using IR sensors, which will directly measure the IR radiation from the crystals. The thermistors will no longer need to be attached to the crystals, but will be mechanically attached to the copper backing.

The next large source of uncertainty is related to the precision of the angular encoders. With a Si (1,1,1) crystal, and a Bragg angle of \unit{\approx 39}{degrees}, the dispersion is such that a \unit{0.2}{arcseconds} accuracy in angle measurement leads to  an uncertainty of \unit{0.0036}{meV} or \unit{1.2}{ppm}.
This would get worse for x~rays of heavier elements,  giving \unit{1.4}{ppm} for the M1 transition in He-like K (\unit{3.47}{keV}), \unit{1.6}{ppm} for the M1 transition in He-like Ca (\unit{3.86}{keV}) and \unit{3.1}{ppm} for He-like Fe  (\unit{6.64}{keV}). Using Si (2,2,0) leads to a very small \unit{0.4}{ppm} uncertainty \textbf{Bragg angle} for the M1 transition He-like potassium. One can obtain \unit{1.7}{ppm} for Fe in first order and \unit{0.2}{ppm} in second order. That measurement would require a very bright x-ray source. One could go beyond this limitation by doing a careful calibration of the encoder using a photoelectronic autocollimator \cite{ldt1984} and a 24-sided optical polygon as has been done at NIST \cite{mlik1992,sdmp1994}.

The last large uncertainty in this category is related to the fact that there are no accurate measurements of the index of refraction of Si at these energies. There has been a proposition to do it by comparing directly the deflection angle in transmission and reflection, but it has not been implemented \cite{hudson2000}.
Such a measurement, if accurate could validate the theoretical or semi-empirical values \cite{hgd1993} (which uses atomic experimental and theoretical photoabsorption cross sections) that we have used and reduce the uncertainty.

\section{Results and discussion}
\label{sec:results}
\begin{figure}
[tb]
\centering
\includegraphics[width=\columnwidth]{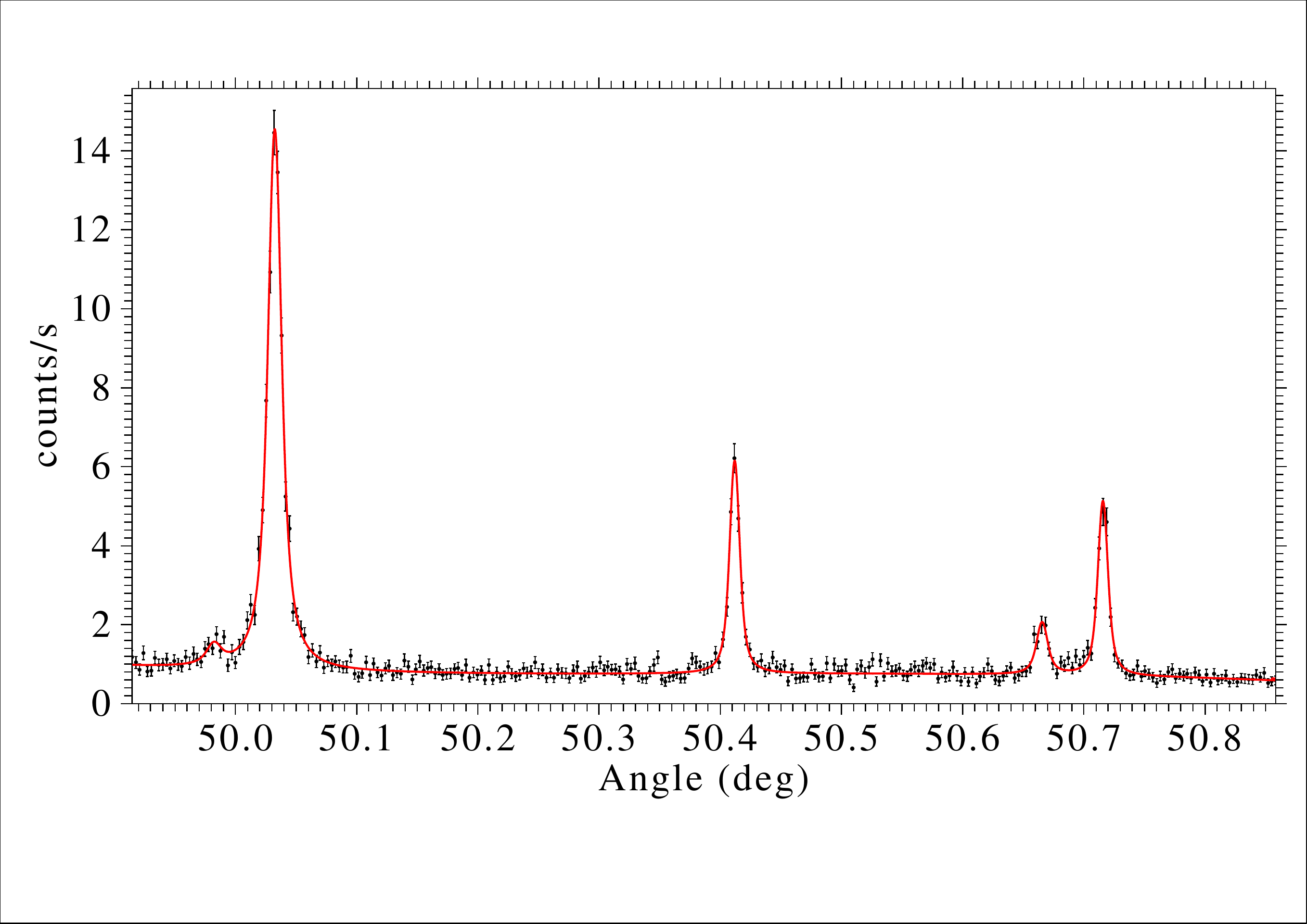}
\caption{(color online) Experimental survey dispersive spectrum. Left peak: $1s^2 2s^2 2p\,^1P_1 \rightarrow 1s^22s^2\,^1S_0$ in Ar$^{14+}$, central peak: $1s2s\,^3 S_1\rightarrow 1s^2\,^1 S_0$ in Ar$^{16+}$, right peaks: $1s2s2p\,^2P_{1/2,\,3/2}\rightarrow1s^22s\,^2S_{1/2}$ in Ar$^{15+}$. The line represents 
a fit using a sum of Voigt profiles.}
\label{fig:exp_Ant}%
\end{figure}

In Figs. \ref{fig:parallel-fit-comp} and \ref{fig:dispersive-fit-comp} we  present
a measurement of the non-dispersive and dispersive spectra obtained with
the DCS for the relativistic M1 transition $1s2s\,^3 S_1\rightarrow 1s^2\,^1 S_0$ in Ar$^{16+}$. The data were acquired by summing individual
back-and-forth 100 bins scans, lasting roughly 10 minutes in the non-dispersive
case and 20 minutes in the dispersive case. The non-dispersive spectrum was acquired in \unit{943}{s} and  the dispersive one in \unit{18240}{s} (these values are corrected for dead time, corresponding to periods when the first crystal position has drifted and is being corrected) 
In Fig. \ref{fig:exp_Ant} we show a survey spectrum, in which the angular range was chosen to includes
peaks corresponding to transition energies of Ar$^{14+}$, Ar$^{15+}$ and Ar$^{16+}$ ions. 
The tallest peak on the left side  corresponds to the transition $1s^2 2s^2 2p\,^1P_1 \rightarrow 1s^22s^2\,^1S_0$ in Ar$^{14+}$.The 
the central peak is the  M1 transition in Ar$^{16+}$. 
The double peak on the right corresponds to the doublet $1s2s2p\,^2P_{1/2,\,3/2}\rightarrow1s^22s\,^2S_{1/2}$ in Ar$^{15+}$.
A description of the mechanism leading to this spectrum can be found in Refs.  \cite{mcsi2001,cmps2001,smci2008,mmcs2009,scmm2010,smcm2011}.

\begin{figure}
[tb]
\centering
\includegraphics[width=\columnwidth]{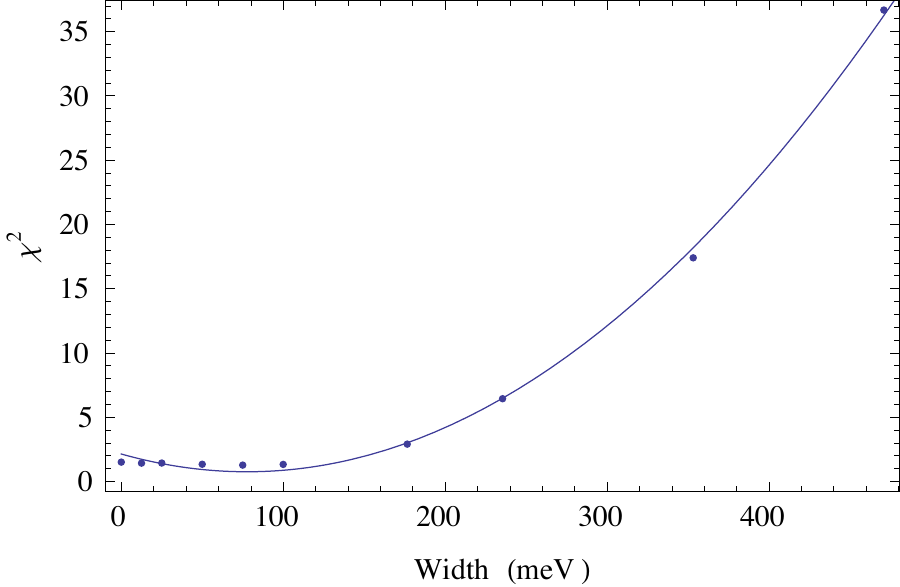}
\caption{Variation of the $\chi^2$ as a function of the Gaussian width introduced in the simulation to represent the Doppler broadening corresponding to the ions temperature.}
\label{fig:chi2-width}%
\end{figure}

The  magnetic dipole (M1) transition has a natural width several orders of magnitude smaller than any line ever measured with a DCS until now. Its measured dispersive width is close to the non-dispersive peak width, which represents the intrinsic response function of the instrument.
The continuous lines in Figs.\ref{fig:parallel-fit-comp} and \ref{fig:dispersive-fit-comp} (right) correspond to simulated profiles fitted to the measured spectra. These simulated profiles were evaluated for the case of an aligned DCS, unpolarized x~rays and a diffraction profile of an ideal flat crystal, with  negligible mosaicity and asymmetric cut. The simulation reproduces the non-dispersive data with a reduced $\chi^2\approx 1.2$. 
This precise fit of the simulated profile on the experimental spectra with no adjustable parameters except line position and intensity validates the hypothesis of perfect crystals and of an ideal alignment of the spectrometer components as made in the simulation.
On the dispersive side, we fitted using simulations with various values of the Gaussian width representing the Doppler broadening due to the ion motion in the ECRIS. The dependance of the $\chi^2$ on the width, corresponding to the spectrum of Fig. \ref{fig:dispersive-fit-comp} is plotted on Fig. \ref{fig:chi2-width}.
The minimum corresponds to a width of \unit{77.6(6.7)}{meV} and to a reduced $\chi^2=0.75$.


\begin{figure}
[tb]
\centering
\includegraphics[clip=true,width=9cm]{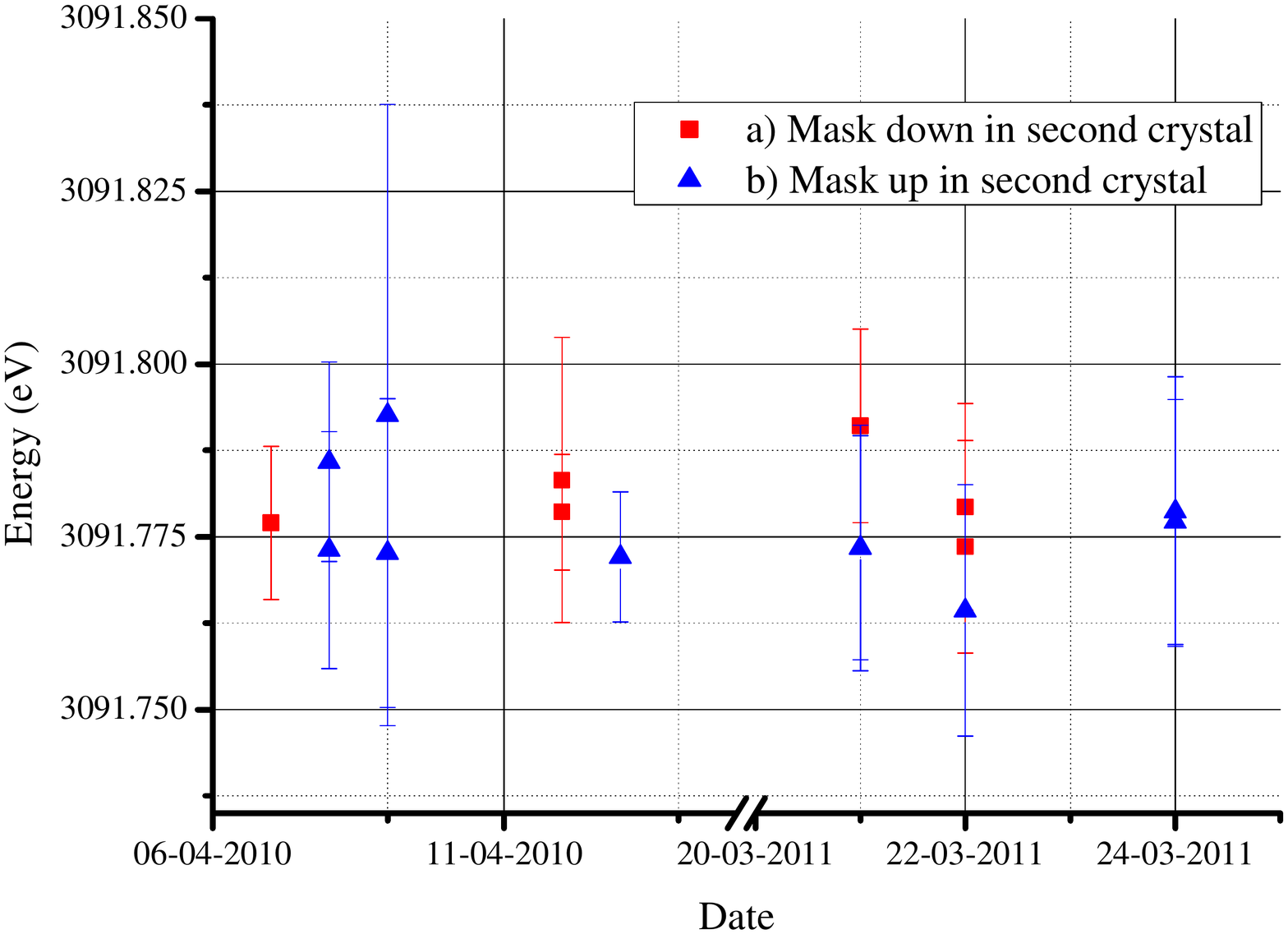}
\caption{ (color online)Plot
of the energy obtained for the mask test. The red squares a) correspond
to the mask placed on the lower half of the second crystal in the DCS. The blue
triangles b) show measurement results for the mask placed on the upper
half of the second crystal.}
\label{fig:dif_ene_masks_verMis}%
\end{figure}

In Sec. \ref{sec:theoryDCS} we discussed  a
method for probing vertical alignment errors using crystal masking. We used this method, performing several measurements using
the Be-like line (the most intense peak in Fig. \ref{fig:exp_Ant}) with a first crystal angle of
\unit{130}{degrees}.  In the first set of measurements, we have
placed a thick brass mask on the upper
half of the second crystal. In the second set of measurements, the lower half
of the second crystal was blocked with the same mask.
Fig. \ref{fig:dif_ene_masks_verMis} shows the line energies obtained
by analyzing all  the measurements performed with either mask positions.
A first set of measurements was performed in April 2010, while
a second set of measurements  was done in March 2011. The
energy obtained in the analysis using Voigt profiles
for mask covering the upper half is \unit{3091.780\pm0.005}{eV}. For a mask covering the upper half of the crystal, it is
\unit{3091.777\pm0.005}{eV}, corresponding to an energy shift of \unit{3\pm7}{meV}. The  uncertainty is only due to statistics. The observed shift is consistent with the one expected from the alignment uncertainty, which is \unit{1.5}{meV} for \unit{0.01}{degrees} as can be seen from Fig. \ref{fig:dif_energ_masks_verMis}. 

\begin{figure}
[tb]
\centering
\includegraphics[clip=true,width= 1.1\columnwidth]{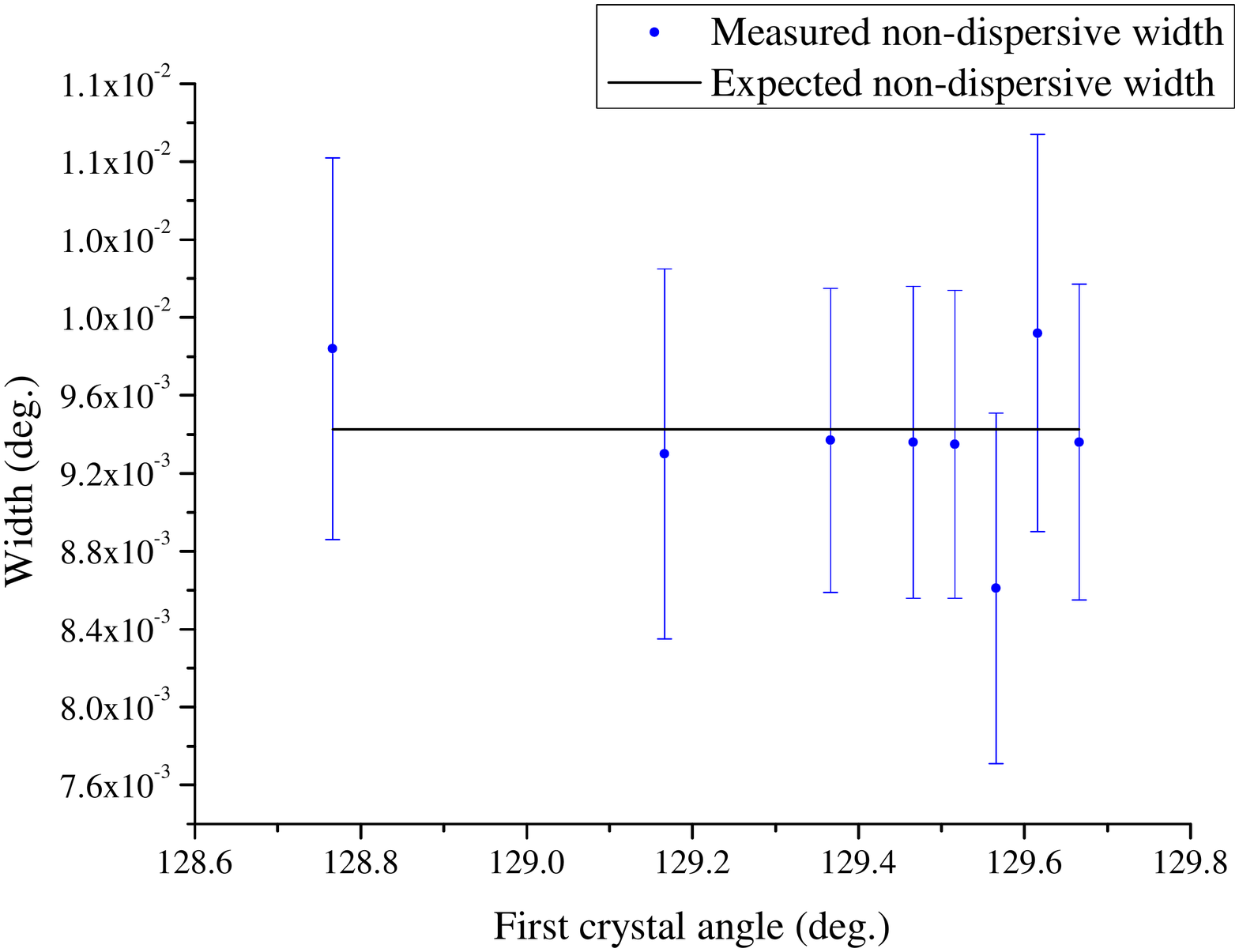}
\caption{(Color online) Measurements of the non-dispersive width for several values of first crystal angle.}%
\label{fig:first_widths_exp}%
\end{figure}

We also experimentally checked if a curvature in both crystals can be neglected. For that matter we performed measurements of the non-dispersive width for several values of the first crystal as is plotted in Fig.~\ref{fig:first_widths_exp}. Comparing Figs. \ref{fig:first_widths} and \ref{fig:first_widths_exp}  we notice that there is no observable dependence  of the width on the first crystal angle within the statistical uncertainty. 

\section{Conclusions}
\label{sec:concl}

We provide a complete description of an experimental set-up composed of a double crystal spectrometer and of an electron-cyclotron resonance ion source, designed to measure low energy x~rays from middle-$Z$ highly-charged ions on an absolute energy scale.
We experimentally demonstrated that the ECRIS plasma yields the necessary x-ray intensity to perform accurate
measurements with a DCS. 
An \emph{ab initio} simulation of the experimental setup (based on the spectrometer
and the source geometry) is presented. The simulations describe very accurately experimental line shapes without adjustable parameters. We show by a complete sequence of measurements and simulations that we understand the
systematic errors within the present statistical accuracy of the experimental spectra. 
 The  spectra presented in this work
clearly show that even a relatively small, permanent magnet ECRIS provides high enough intensities for
precision measurements of transitions in highly charged ions with a DCS. 
We also show that our understanding of the line shape is such that we  can investigate the ion temperatures in the plasma.  We are thus now able to obtain values of the natural line  widths in ions with 2, 3 or 4 electrons, leading to a better understanding of the Auger and radiative contributions to the width.

The world-wide unique combination of the DCS and the ECRIS allows to perform high-precision, reference-free
measurements of x-ray transition energies in highly charged ions. These high precision measurements enable direct tests of  QED and many-body effects in middle-$Z$ elements and
will provide new x-ray standards based on narrow transitions of highly charged ions. A \unit{2.5}{ppm} measurement of the $1s2s\,^3 S_1\rightarrow 1s^2\,^1 S_0$  transition energy in Ar$^{16+}$ obtained with this set-up has been published recently\cite{asgl2012}.
%


\section*{acknowledgments}
\label{sec:ack}
Laboratoire Kastler Brossel is ``UMR n$^{\circ}$ 8552'' of the ENS, CNRS and UPMC.
The SIMPA ECRIS has been financed by grants from CNRS, MESR,  and  UPMC. The experiment is supported by grants from BNM \emph{01 3 0002}, the ANR 
\emph{ANR-06-BLAN-0223} and  the Helmholtz Alliance \emph{HA216/EMMI}.

This research was also supported in part by FCT (PEst-OE/FIS/UI0303/2011, Centro de F\'isica At\'omica), by the French-Portuguese collaboration (PESSOA Program, contract no 441.00), by the Ac\c{c}\~oes Integradas Luso-Francesas (contract no F-11/09), and by the Programme Hubert Curien PESSOA 20022VB.
M. G. acknowledges the support of the FCT, under Contract SFRH/BD/38691/2007.
P. A. acknowledges the support of FCT, under Contract No. SFRH/BD/37404/2007 and the German Research Foundation (DFG) within the Emmy Noether program under Contract No. TA 740 1-1.
We thank J.P.~Okpisz, B. Delamour, M. Boujrad, A.Vogt, and S. Souramassing for technical
support, B. Manil for his help at the beginning of the spectrometer construction,  and the ASUR team from INSP. 
We thank L. Hudson, J. Gillaspy and T. Jach for helpful discussions. 

%
%
\bibliographystyle{model5-names}
\bibliography{refs-long}


\end{document}